\documentclass[useAMS,usenatbib]{mnras}

\usepackage{txfonts,graphicx,amssymb,rotating}

\usepackage{calligra}
\DeclareMathAlphabet{\mathcalligra}{T1}{calligra}{m}{n}
\DeclareFontShape{T1}{calligra}{m}{n}{<->s*[2.2]callig15}{}

\usepackage[cal=boondoxo]{mathalfa}

\title[Magnetospheres of hot Jupiters]{Magnetospheres of hot Jupiters: hydrodynamic models \& ultraviolet absorption}

\author[Alexander et al.]{R.D.Alexander$^{1,}$\thanks{email: richard.alexander@leicester.ac.uk}, G.A.Wynn$^1$, H.Mohammed$^{1,2}$, J.D.Nichols$^1$ and B.Ercolano$^{3,4}$ \\$^1$Department of Physics \& Astronomy, University of Leicester, Leicester, LE1 7RH, UK\\$^2$University of Sulaimani, Raparin Way, Sulaimani, Kurdistan Region of Iraq\\$^3$University Observatory, Ludwig-Maximilians-University Munich, Scheinerstr.\,1, D-81679 Munich, Germany\\$^4$Excellence Cluster Universe, Boltzmannstr.\,2, D-85748 Garching, Germany}
  
\begin{document}
\voffset=-0.25in
\def\apj{ApJ}
\def\mnras{MNRAS}
\def\araa{ARA\&A}                
\def\aap{A\&A}                   
\def\aaps{A\&AS}          
\def\apjs{ApJS}               
\def\pasp{PASP}                 
\def\apjl{ApJ}                 
\def\apjs{ApJS}                 
\def\pasj{PASJ}
\def\aj{AJ}
\def\nat{Nat}
\def\sci{Sci}
\def\aapr{A\&AR} 

\newcommand{\Msunyr}{M$_{\odot}$\,yr$^{-1}$}
\newcommand{\Msun}{M$_{\odot}$}
\newcommand{\Mjup}{M$_{\mathrm {Jup}}$}
\newcommand{\ap}{$a_{\mathrm p}$}
\newcommand{\Rp}{$R_{\mathrm p}$}
\newcommand{\Rsun}{R$_{\odot}$}
\newcommand{\Rjup}{R$_{\mathrm {Jup}}$}
\newcommand{\kms}{km\,s$^{-1}$}

\newcommand{\scripty}[1]{\ensuremath{\mathcalligra{#1}}}

\pagerange{\pageref{firstpage}--\pageref{lastpage}} \pubyear{2015}

\date{Accepted 2015 December 4. Received 2015 December 4; in original form 2015 October 2}

\maketitle

\label{firstpage}

\begin{abstract}
We present hydrodynamic simulations of stellar wind--magnetosphere interactions in hot Jupiters such as WASP-12b.  For fiducial stellar wind rates we find that a planetary magnetic field of a few G produces a large magnetospheric cavity, which is typically 6--9 planetary radii in size.  A bow shock invariably forms ahead of the magnetosphere, but the pre-shock gas is only mildly supersonic (with typical Mach numbers of $\simeq 1.6$--1.8) so the shock is weak.  This results in a characteristic signature in the ultraviolet light curve: a broad absorption feature that leads the optical transit by 10--20\% in orbital phase.  The shapes of our synthetic light-curves are consistent with existing observations of WASP-12b, but the required near-UV optical depth ($\tau \sim 0.1$) can only be achieved if the shocked gas cools rapidly. We further show that radiative cooling is inefficient, so we deem it unlikely that a magnetospheric bow shock is responsible for the observed near-UV absorption. Finally, we apply our model to two other well-studied hot Jupiters (WASP-18b and HD209458b), and suggest that UV observations of more massive short-period planets (such as WASP-18b) will provide a straightforward test to distinguish between different models of circumplanetary absorption.
\end{abstract}



\section{Introduction}
Of the thousands of known exoplanets, the ``hot Jupiters'' are perhaps the least akin to anything we see in the Solar System.  These giant planets orbit their parent stars in just a few days, and consequently are subject to a variety of extreme physical processes that do not affect other planets.  For example, hot Jupiters are subject to strong tidal forces, which dissipate orbital eccentricity and may inflate the planets' atmospheres \citep*{rasio96,bodenheimer01}.  In many cases hot Jupiters are also inflated by stellar irradiation \citep{sg02}, and their atmospheres can become so extended that they overflow the planets' Roche lobes \citep*{gu03}.  Their short orbital periods also make hot Jupiters, particularly those which transit their host stars, highly amenable to follow-up observations, providing us with a unique laboratory in which to test our understanding of planetary physics.

Among the hot Jupiters discovered to date, WASP-12b \citep{hebb09} is perhaps the most interesting.  With a period of just 1.09d it is one of the shortest-period planets known, and consequently it is one of only a handful of exoplanet systems to have been observed in and out of transit in multiple wavebands.  Ultraviolet (UV) observations with the {\it Hubble Space Telescope (HST)} have found evidence for a deeper and broader transit in the near-UV than is seen at optical or infrared wavelengths \citep{fossati10,haswell12,nichols15}, which is strongly suggestive of absorbing material around the planet.  Several models have been suggested to explain this excess absorption, most notably Roche lobe overflow from the planet \citep{lai10,bisikalo13} and a bow shock formed where the stellar wind and/or corona interacts with the planetary magnetic field \citep{lai10,vidotto10,llama11}.  However, to date these models have been somewhat idealised, and as a result current data do not allow us to distinguish between different scenarios.

Here we present detailed hydrodynamic models of the interaction between stellar winds and the planetary magnetospheres of hot Jupiters.  We work within the existing picture of absorption in a magnetospheric bow shock \citep{lai10,vidotto10,llama11} but, for the first time, build a self-consistent hydrodynamic model of both the stellar wind and the shock.  We find that the bow shock is always weak, and the shock structure differs substantially from that assumed by \citet{vidotto10} and \citet{llama11}. We then use our hydrodynamic models to compute theoretical UV light-curves, and investigate how the UV transit shape varies with different physical model parameters (such as the wind temperature and planetary magnetic field strength). The shape of our model transits is consistent with existing UV data for WASP-12b, but additional radiative transfer calculations suggest that the bow shock does not have sufficient opacity at (near-)UV wavelengths to reproduce the observations (unless the shocked gas is able to cool very rapidly). We also show that 100\% phase coverage is highly desirable if we are to use such observations to test theoretical models in detail. Finally we construct models of other known systems, and show that additional UV observations of more massive short-period hot Jupiters \citep[such as WASP-18b;][]{hellier09} should distinguish clearly between different models for circumplanetary absorption.


\section{Model}
We use the {\sc zeus-2d} hydrodynamics code \citep{sn92} to simulate the interaction between a stellar wind and a planetary magnetosphere.  We work in two dimensions, using a polar ($r,\phi$) grid with a volume element which scales as $\Delta(r^3/3)$.  The computational grid spans the range [0,2$\pi$] in the $\phi$-direction, with periodic boundary conditions; we effectively consider an infinitesimal midplane ``wedge'' of a 3-D spherical polar grid\footnote{The {\sc zeus-2d} algorithms are formulated in a manner that allows the user to adopt any set of orthogonal basis vectors, by specifying the appropriate metric coefficients $\{h_1,h_2,h_3\}$. Here these coefficients are $h_1 = 1$, $h_2 = r\sin\phi$ and $h_3 = r$.}.  The grid cells are logarithmically-spaced in $r$ and linearly-spaced in $\phi$, with the numbers of radial ($N_r$) and azimuthal ($N_{\phi}$) grid cells chosen so that the grid cells are approximately square (i.e., $\Delta r = r \Delta \phi$).  We adopt the van Leer (second order) interpolation scheme and the standard von Neumann \& Richtmyer form for the artificial viscosity (with $q_{\mathrm {visc}} = 2.0$); tests indicate that neither of these choices has a significant influence on our results.  We adopt a system of units such that the unit of mass is the stellar mass $M_*$, the unit of length is the planet semi-major axis \ap, and the unit of time is the planet's orbital period $P$.  This sets the gravitational constant $G=4\pi^2$ in code units, and the planet's (Keplerian) orbital velocity $u_{\mathrm p} = \sqrt{GM_*/a_{\mathrm p}} = 2\pi$.  


\subsection{Stellar Wind}\label{sec:wind}
\begin{figure}
\centering
       \resizebox{\hsize}{!}{
       \includegraphics[angle=270]{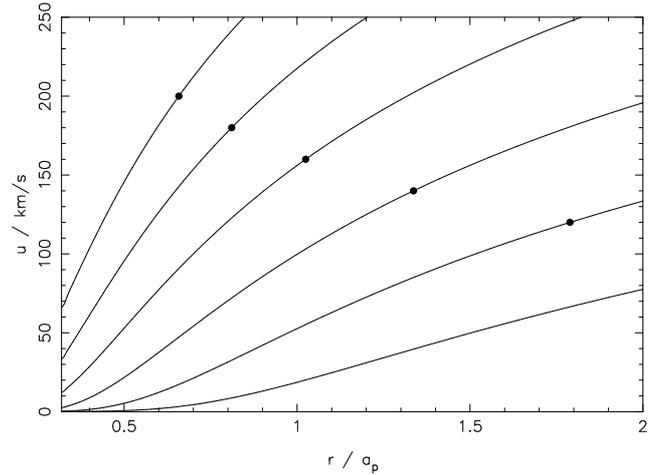}
       }
       \caption{Velocity profiles for isothermal (Parker) winds, calculated using {\sc zeus-2d} and scaled to the parameters of the WASP-12 system ($M_* = 1.35$\Msun, $a_{\mathrm p} = 0.0229$AU).  Profiles, from bottom to top respectively, are plotted for sound speeds $c_{\mathrm s}$=100, 120, 140, 160, 180 \& 200\kms.  In each case the filled black circle denotes the location of the sonic point.}     
           \label{fig:parker_vel}
\end{figure}
\begin{figure}
\centering
       \resizebox{\hsize}{!}{
       \includegraphics[angle=270]{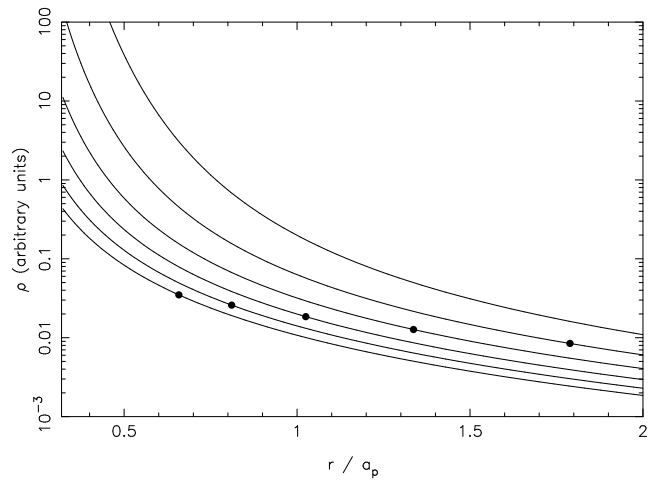}
       }
       \caption{Density profiles for the (Parker) wind solutions in Fig.\,\ref{fig:parker_vel}.  The absolute density in the Parker solution is scale-free, so we adopt arbitrary units on the vertical axis and scale our solutions to a constant wind rate $\dot{M}_{\mathrm w}=4\pi r^2 u \rho$.  As in Fig.\,\ref{fig:parker_vel} the profiles, from top to bottom respectively, are plotted for sound speeds $c_{\mathrm s}$=100, 120, 140, 160, 180 \& 200\kms, with the sonic points marked by circles.  Note that a relatively small (factor of two) decrease in the sound speed dramatically increases the density gradient $d(\log\rho)/dr$, and consequently increases the density contrast between the inner boundary (the stellar surface) and the planet's position by more than three orders of magnitude.}     
           \label{fig:parker_rho}
\end{figure}

We model the stellar wind as a spherically-symmetric isothermal (Parker) wind.  This problem has a well-known analytic solution \citep{parker58,cranmer04}, which is fully specified by the choice of sound speed $c_{\mathrm s}$ and passes through a sonic transition at $r_{\mathrm s} = GM_*/2c_{\mathrm s}^2$.  We implement this in {\sc zeus-2d} by specifying an ``inflow'' inner boundary condition at the stellar radius $r_{\mathrm {in}} = R_*$, with the injection velocity $u_{\mathrm {in}}$ equal to that of the Parker solution at that radius [i.e., $u_{\mathrm {in}} \equiv u_{\mathrm w}(r_{\mathrm {in}})$].  The gas density appears in the Parker solution only as a normalisation constant, and consequently the density in our simulations is arbitrary; we normalise to the (fixed) value at the inner boundary, $\rho_0$.  We scale our fiducial model to physical units by adopting the parameters of the WASP-12b system (see Section \ref{sec:wasp12} below) and therefore set $R_*$=0.319\ap.   We place the outer boundary at 10 times the planet's orbital radius, so our grid spans $[r_{\mathrm {in}},r_{\mathrm {out}}]=[0.319a_{\mathrm p},10.0a_{\mathrm p}]$ in the radial direction.  Our standard models have $N_{\phi} = 250$ grid cells in the azimuthal direction, and therefore $N_{\mathrm r} = 139$ cells in the radial direction.  We initially set the density everywhere on the grid to a small value ($10^{-15}\rho_0$), and allow the wind to evolve to a steady state.  Slow stellar winds from Sun-like stars typically have temperatures $\sim$1MK \citep[e.g.,][]{lc99}, so we explore a range of sound speeds from 100--200\kms.  This approach reproduces the Parker wind solution to high accuracy, matching the analytic solution to within approximately 0.1\%.  Example velocity and density profiles (for a fixed wind rate $\dot{M}_{\mathrm w}$ and different choices of the sound speed $c_{\mathrm s}$) are shown in Figs.\,\ref{fig:parker_vel} \& \ref{fig:parker_rho}.  

\subsection{Planet}\label{sec:planet}
We wish to model the interaction of a close-in planet with the stellar wind, by adding both the planet's gravity and the effect of its magnetic field.  We assume a fixed, circular orbit for the planet, which has mass $M_{\mathrm p}$, semi-major axis $a_{\mathrm p}$ and orbital frequency $\Omega$.  The planet's position at time $t$ is therefore $\mathbf r_{\mathrm p} = (r_{\mathrm p},\phi_{\mathrm p}) = (a_{\mathrm p},\Omega t)$.

{\sc zeus-2d} solves the momentum equation in the form
\begin{equation}
\rho \frac{d\mathbf u}{dt}+\mathbf u.\nabla\mathbf u = -\nabla p +\rho\mathbf g \, .
\end{equation}
In the absence of a planet the gravitational acceleration is simply that due to the star, so $\mathbf g = -(GM_*/r^2)\hat{\mathbf r}$.  We neglect self-gravity, radiation hydrodynamics and magnetic fields in the gas (these modules are switched off in the code), and incorporate accelerations due to the planet by adding an additional term $\rho\mathbf a$ to the right-hand-side of the momentum equation.  The gravitational acceleration $\mathbf g$ appears explicitly in the {\sc zeus-2d} code as a source term, and we compute the accelerations due to the planet $\mathbf a$ in a similar manner.

\begin{figure*}
\centering
       \resizebox{\hsize}{!}{
       \includegraphics[angle=270]{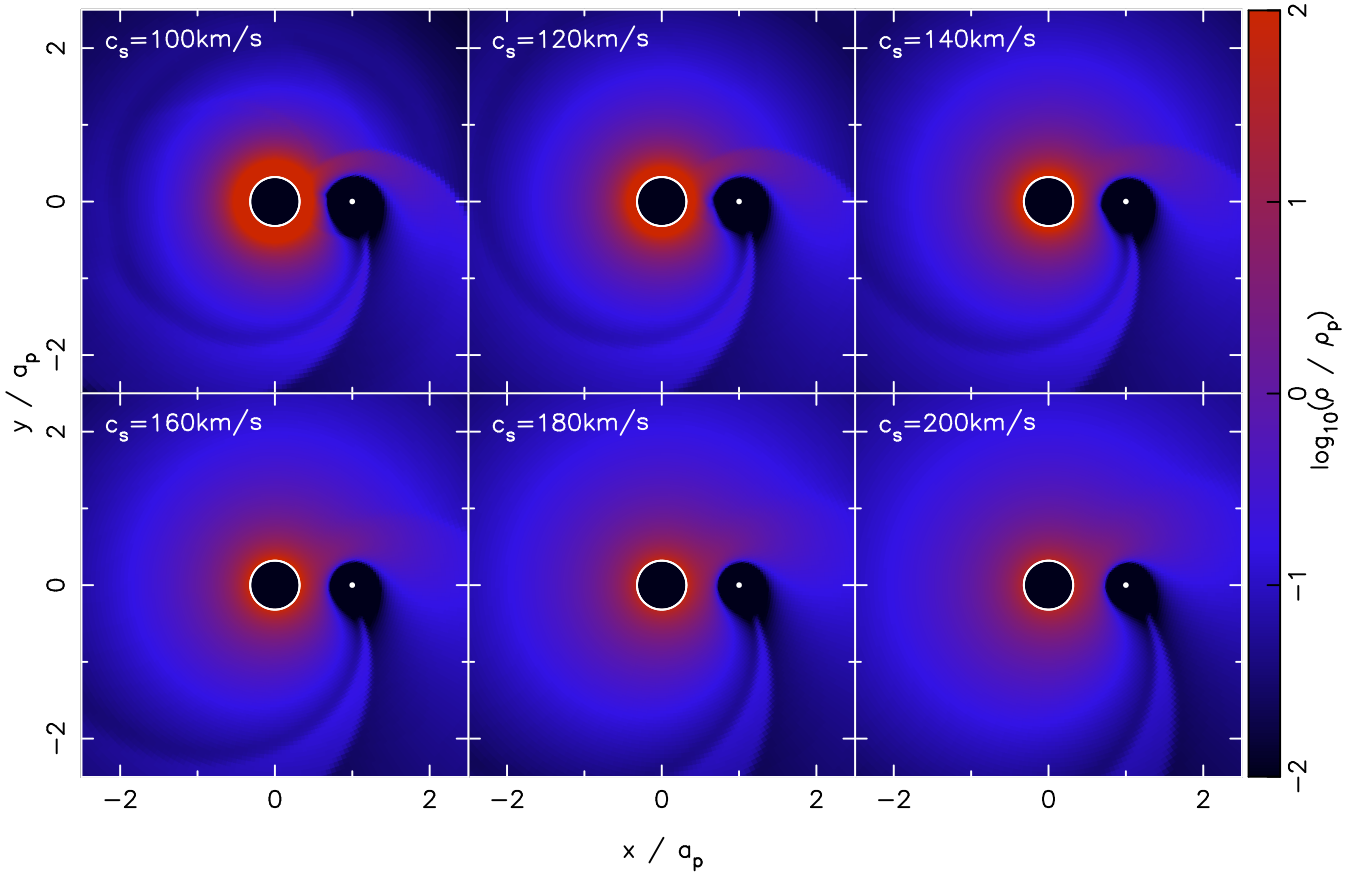}
       }
       \caption{Steady-state density structure in our simulations with $C_{\mathrm B} = 0.3$.  The white line denotes the inner grid boundary at $r_{\mathrm {in}}$$=$$R_*$, while the filled white circle denotes the position and radius of the planet (i.e., both the star and planet are plotted to scale).  In each panel the density is normalised to the value at the planet's orbital radius, 180$^{\circ}$ out of phase with the planet (denoted by $\rho_{\mathrm p}$).  The magnetospheric radius is approximately constant in all the models, but as the sound speed increases the cavity and wake are angled progressively more towards the star.  In addition, lower sound speeds result in slightly stronger (though still weak) bow shocks, with more pronounced density enhancements ahead of the planet.}
           \label{fig:density_maps}
\end{figure*}

We assume that the planet has a dipolar magnetic field aligned with the planet's orbital angular momentum, and work in the far-field limit (i.e., we assume $\scripty{r} = |\mathbf r-\mathbf r_{\mathrm p}| \gg R_{\mathrm p}$, where $R_{\mathrm p}$ is the radius of the planet), so the magnetic field strength scales as $B \propto \scripty{r}^{-3}$. The magnetic pressure $P_{\mathrm B} = B^2/8\pi$, and the acceleration due to the magnetic pressure is $(-1/\rho) (\partial P_{\mathrm B}/\partial \scripty{r}) \propto \scripty{r}^{-7}$.  The acceleration due to the planet at an arbitrary position $\mathbf r$ is therefore
\begin{equation}\label{eq:a}
\mathbf a = \left( \frac{C_{\mathrm B}}{|\mathbf r-\mathbf r_{\mathrm p}|^8} - \frac{GM_{\mathrm p}}{|\mathbf r-\mathbf r_{\mathrm p}|^3} \right) (\mathbf r-\mathbf r_{\mathrm p}) \, .
\end{equation}
where $C_{\mathrm B}$ is a normalisation constant that sets the magnitude of the ``magnetic'' accelerations.  We adopt this formalism in order to preserve the scale-free behaviour of the Parker wind: the additional accelerations due to the planet depend only on $M_{\mathrm p}$, $C_{\mathrm B}$, and position, and are independent of the density normalisation.  We treat $C_{\mathrm B}$ as an input parameter, which effectively determines the magnetospheric radius (see Section \ref{sec:sims} and Figs.\,\ref{fig:density_orbit_cs} \& \ref{fig:density_orbit_CB}).  However, the magnetic field strength is not scale-free, as the relationship between $C_{\mathrm B}$ and $B$ depends on the local gas density $\rho$.  If we follow convention and specify the planetary magnetic field strength in terms of the surface field $B_0$, then Equation \ref{eq:a} implies that $C_{\mathrm B} = 3 B_0^2 R_{\mathrm p}^6 / 4 \pi \rho$.  

Equation \ref{eq:a} diverges as $|\mathbf r-\mathbf r_{\mathrm p}| \rightarrow 0$, so we soften the potential in order to prevent numerical errors close to the planet's position.  We compute the accelerations as
\begin{equation}
\mathbf a = (a_{\mathrm B} + a_{\mathrm g})(\mathbf r-\mathbf r_{\mathrm p}) \, ,
\end{equation}
where $a_{\mathrm B}$ and $a_{\mathrm g}$ are the magnitudes of the magnetic and gravitational terms, respectively.  We use Plummer softening for the gravitational potential
\begin{equation}
a_{\mathrm g} = - \frac{GM_{\mathrm p}}{[\scripty{r}^2 + (\epsilon R_{\mathrm p})^2]^{3/2}}
\end{equation}
and soften the magnetic potential using a Gaussian function 
\begin{equation}
a_{\mathrm B} = \left\{ \begin{array}{ll}
0 & \textrm{if } \, \scripty{r} < \delta R_{\mathrm p}\\
\frac{C_{\mathrm B}}{\scripty{r}^8} \exp\left(-\frac{[\scripty{r}-(1+\delta) R_{\mathrm p}]^2}{2 (\delta \mathrm R_{\mathrm p})^2}   \right) & \textrm{if } \, \delta R_{\mathrm p} \le  \scripty{r} \le (1+\delta) R_{\mathrm p} \\
\frac{C_{\mathrm B}}{\scripty{r}^8} & \textrm{if } \, \scripty{r} > (1+\delta) R_{\mathrm p} \, .\\
\end{array}\right.
\end{equation}
We adopt softening parameters $\epsilon = 0.4$ and $\delta = 0.15$ throughout.  In practice the softened potential deviates from the true potential only for $\scripty{r} < 2$\Rp\ (i.e., within one planetary radius of the planet's surface), where the gas density is very low, and tests indicate that modest variations in these softening parameters have no significant effect on our results.  We then decompose the vector $\mathbf a$ into its components $(a_r,a_{\phi})$ (see Appendix \ref{sec:appendix}) and add these these terms as explicit accelerations in the {\sc zeus-2d} source step.  We also impose a numerical density floor at $10^{-15}\rho_0$, in order to prevent the low-density region close to the planet restricting the time-step to unreasonably small values.  We have parallelised the {\sc zeus-2d} code, for a shared-memory architecture, using the OpenMP formalism\footnote{See {\tt http://openmp.org}}.  Our simulations were run on the ALICE\footnote{See {\tt http://go.le.ac.uk/alice}} and DiRAC2/{\it Complexity}\footnote{See {\tt http://www.dirac.ac.uk}} high-performance computing clusters at the University of Leicester.


\section{WASP-12b models}\label{sec:wasp12}
For our fiducial model set we scale our simulations to the WASP-12b system.  WASP-12b is the prototype short-period hot Jupiter, and consists of an approximately Jupiter-mass planet in a 1.09d orbit around a Sun-like (late-F / early-G type) star \citep{hebb09}.  It is also the best-studied of the handful of exoplanets which have been observed in transit at near-UV wavelengths \citep{fossati10,haswell12,nichols15}.  We adopt the parameters of the WASP-12b system derived by \citet{hebb09}: $M_* = 1.35$\Msun, $a_{\mathrm p} = 0.0229$AU, $P = 1.09$d (and therefore $u_{\mathrm p} = 228.7$\kms), $R_*$=1.57\Rsun(=0.319\ap), $M_{\mathrm p}$=1.41\Mjup($=$$9.97\times10^{-4}M_*$) and $R_{\mathrm p}$=1.79\Rjup($=$0.0365\ap).

We note that with these parameters the stellar wind velocity relative to the planet $u = \sqrt{u_{\mathrm p}^2 + u_{\mathrm w}^2(a_{\mathrm p})}$ is always supersonic.  For low sound speeds the orbital velocity $u_{\mathrm p}$ is supersonic, while for high sound speeds ($c_{\mathrm s}>u_{\mathrm p}$) the sonic radius $r_{\mathrm s} < a_{\mathrm p}$, so the wind is supersonic by the time it reaches the planet's orbit.  In the frame of the planet the minimum Mach number at $r=a_{\mathrm p}$ is in fact $\mathcal M =1.73$, which occurs for $c_{\mathrm s}$=161\kms.  However, in all cases of interest the wind is only modestly supersonic: $\mathcal M<2.3$ for $c_{\mathrm s}$=100--300\kms.  We therefore expect any interactions between the planet and the stellar wind in the WASP-12b system to result in weak shocks.


\subsection{Simulations}\label{sec:sims}
\begin{figure}
\centering
       \resizebox{\hsize}{!}{
       \includegraphics[angle=270]{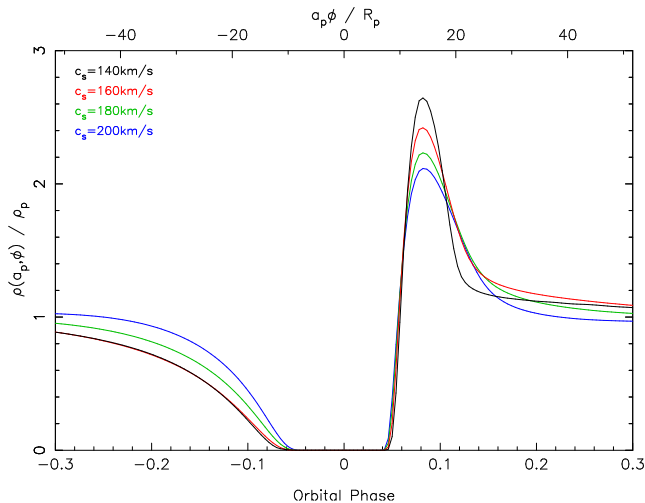}
       }
       \caption{Azimuthal density profiles at the planet's orbital radius, for models with $C_{\mathrm B}$=0.3 and various value of the sound speed $c_{\mathrm s}$.  (For clarity, we omit the models with $c_{\mathrm s}$=100 \& 120\kms.)  As in Fig.\,\ref{fig:density_maps}, the density is normalised to the value 180$^{\circ}$ out of phase with the planet.  The lower axis shows the azimuthal coordinate in units of orbital phase (which ranges from $[-0.5,0.5]$), while the upper axis shows the distance around the planet's orbit in units of the planet radius $R_{\mathrm p}$.  The size of the magnetospheric cavity ($\simeq$7.5\Rp\ in radius) and the position of bow shock (with peak density $\simeq$14\Rp\ ahead of the planet) are essentially independent of the sound speed in the wind, but the density contrast in the shock decreases with increasing $c_{\mathrm s}$.}
           \label{fig:density_orbit_cs}
\end{figure}

\begin{figure}
\centering
       \resizebox{\hsize}{!}{
       \includegraphics[angle=270]{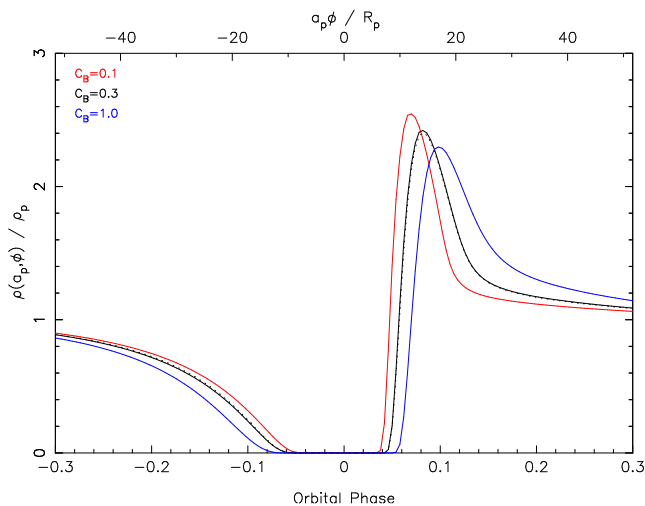}
       }
       \caption{As Fig.\,\ref{fig:density_orbit_cs}, but for models with $c_{\mathrm s}$=160\kms\ and various values of the magnetic constant $C_{\mathrm B}$.  The magnetospheric cavity increases in size for stronger magnetic fields, from $\simeq$6.0\Rp\ for $C_{\mathrm B}$=0.1 to $\simeq$9.0\Rp\ for $C_{\mathrm B}$=1.0. The dashed black line denotes the simulation with $C_{\mathrm B}$=0.3 run at twice the numerical resolution ($N_{\phi}$=500, $N_r$=276), and shows that our simulations have achieved convergence.}
           \label{fig:density_orbit_CB}
\end{figure}

When our hydrodynamic simulations are allowed to evolve the planet carves out a magnetospheric cavity in the stellar wind, and (in the frame of the planet) the simulations rapidly evolve into a steady state.  This steady state is typically reached after 3--5 planetary orbits; we take the results after 10 orbits as the final flow solutions.  We computed models for sound speeds $c_{\mathrm s}$=100, 120, 140, 160, 180 \& 200\kms, and magnetic constants $C_{\mathrm B}$=0.1, 0.3 \& 1.0.  The resulting steady-state density structures (for $C_{\mathrm B}$=0.3) are shown in Fig.\,\ref{fig:density_maps}; azimuthal density profiles $\rho(a_{\mathrm p},\phi)$ are shown in Figs.\,\ref{fig:density_orbit_cs} \& \ref{fig:density_orbit_CB} (for varying sound speed $c_{\mathrm s}$ and magnetic constant $C_{\mathrm B}$, respectively).  Fig.\,\ref{fig:density_orbit_CB} also shows the result of a numerical convergence test run at twice our standard numerical resolution (i.e., $N_{\phi}$=500, $N_r$=276): no significant differences are seen at higher resolution, indicating that our simulations are well-resolved and numerically robust.

In all cases we see qualitatively similar flow solutions.  The planetary magnetic field carves out an large, near-circular magnetospheric cavity, which is preceded by a near semi-circular bow shock and followed by a bifurcated wake.  The geometry of the magnetosphere changes with sound speed, with the cavity and wake oriented progressively more towards the star as the sound speed increases \citep[as suggested by][]{vidotto10}.  For $c_{\mathrm s}$=100\kms\ the bow shock is almost perpendicular to the planet's orbit, but for sound speeds $>$160\kms\ the wind velocity at \ap\ is comparable to the planet's orbital speed (228.7\kms), and the shock is angled significantly towards the star.

As expected, in all our models the bow shock is weak.  The peak Mach number in the pre-shock gas (ahead of the magnetosheath) ranges from $\mathcal M \simeq 1.6$--1.8 for the models with $c_{\mathrm s}$=140--200\kms.  For lower sound speeds we see slightly stronger shocks, but even for $c_{\mathrm s}$=100\kms\ the peak Mach number is only $\mathcal M \simeq 2.3$.  Consequently the shocks are broad: the magnetosheaths have widths comparable to the radius of the magnetospheric cavities, and have only modest density enhancements.  The density contrast between the pre- and post-shock gas exceeds 2.5 only for  $c_{\mathrm s}$$<$140\kms, and the shocks are never strong enough to produce a density discontinuity.  In the models with the lowest sound speeds ($c_{\mathrm s}$=100 \& 120\kms) there is sufficient momentum in the shocked gas that we see some spurious reflection of material from the inner grid boundary.  This is a real physical effect (as our grid boundary corresponds to the photospheric surface of the star), but we make no attempt to model it accurately.  However, these reflections have only a small effect on the flow structure, and essentially no effect on the light-curves computed in Section \ref{sec:lightcurves}.

Our simulations are parametrized such that the size of the magnetosphere is essentially independent of the sound speed in the gas (as can be seen in Figs.\,\ref{fig:density_maps} \& \ref{fig:density_orbit_cs}), and depends only on the magnetic constant $C_{\mathrm B}$.  The choice of magnetic constant therefore specifies the magnetospheric radius, with values $C_{\mathrm B}$=0.1, 0.3 \& 1.0 corresponding to cavity radii of $\simeq$6.0, 7.5 \& 9.0\Rp, respectively (see Fig.\ref{fig:density_orbit_CB}).  In our simulations (which neglect the stellar magnetic field) the magnetospheric radius is determined by the balance between ram pressure in the wind and magnetic pressure from the planet (see Section \ref{sec:planet}), with 
\begin{equation}\label{eq:B_CB}
B_0 = \sqrt{(4/3)\pi\rho C_{\mathrm B}} R_{\mathrm p}^{-3} \, .
\end{equation}
For a fiducial stellar wind rate $\dot{M}_{\mathrm w} = 10^{-15}$\Msunyr\ and $c_{\mathrm s}$=160\kms$^,$\footnote{In the planet's frame, this corresponds to a dynamic pressure at the planet's orbital radius of $P_{\mathrm w}=\rho(u^2 + c_{\mathrm s}^2) \simeq 3\times10^{-6}$g\,cm$^{-1}$\,s$^{-2}$.}, $C_{\mathrm B}$=0.3 therefore equates to a surface magnetic field $B_0 \sim 4$G; we discuss more realistic estimates of the field strength in Section \ref{sec:limitations}.  In all cases the half-width of the magnetosheath (i.e., the distance between the outer edge of the cavity and the peak gas density) is approximately the same as the magnetospheric radius.  The planetary magnetic field therefore carves out a large, extended structure, with the shock and cavity typically stretching around $\gtrsim$30\% of the planet's orbit.

  
\subsection{Light-curves}\label{sec:lightcurves}
\begin{figure}
\centering
       \resizebox{\hsize}{!}{
       \includegraphics[angle=270]{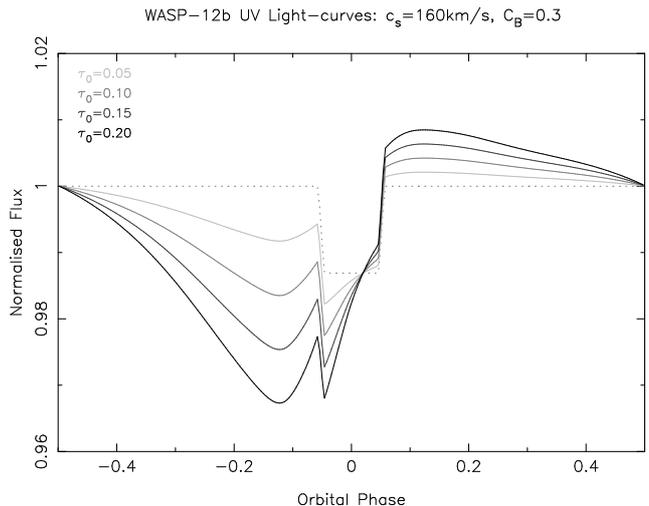}
       }
       \caption{WASP-12b UV light curves for our fiducial model ($c_{\mathrm s}$=160\kms\ and $C_{\mathrm B}$=0.3), calculated for various optical depth parameters $\tau_0$.  For reference, the optical transit $I_{\mathrm p}(\phi)$ is shown as a dotted grey line.  The minimum of the UV transit corresponds to the peak density in the magnetosheath, and leads the planet in orbital phase by $\simeq$12--13\%.  Note also that lines-of-sight through the magnetospheric cavity have relatively low column densities, resulting in a normalised flux greater than unity in the post-transit region.}
           \label{fig:light_curves_tau}
\end{figure}
Having found steady-state flow solutions, we now compute UV light-curves for the combined transit of the planet and magnetosphere.  As a first approximation we assume a constant mass absorption coefficient (opacity) in the wind. This approximation is exact in the limit of constant density and temperature, and as the near-UV opacity in the wind is dominated by metal lines it varies only weakly with density. Constant opacity is therefore a reasonable first assumption for our (isothermal) simulations; we estimate the magnitude of the opacity and discuss the validity of this approximation further in Sections \ref{sec:mocassin} \& \ref{sec:dis}.

We calculate the gas column density $\Sigma(\phi)$ from our steady-state solutions by integrating the gas density along the line-of-sight to the star.  Using the density structures from our 2-D simulations, we compute the absorbing column at an arbitrary angle $\phi_0$ as
\begin{equation}
\Sigma(\phi_0) = \frac{1}{2R_*} \int_{y=-R_*}^{+R_*} \int_{x=0}^{r_{\mathrm {out}}} \rho(x,y) \, dx\,dy \,\, ,
\end{equation}
where $x = r\cos(\phi-\phi_0)$ and $y = r\sin(\phi-\phi_0)$.  This procedure essentially rotates our steady-state flow structure to compute the absorbing column around the orbit, and by considering values of $\phi_0$ spanning the range $[-\pi,\pi]$ we can generate synthetic light-curves with 100\% phase coverage.

\begin{figure}
\centering
       \resizebox{\hsize}{!}{
       \includegraphics[angle=270]{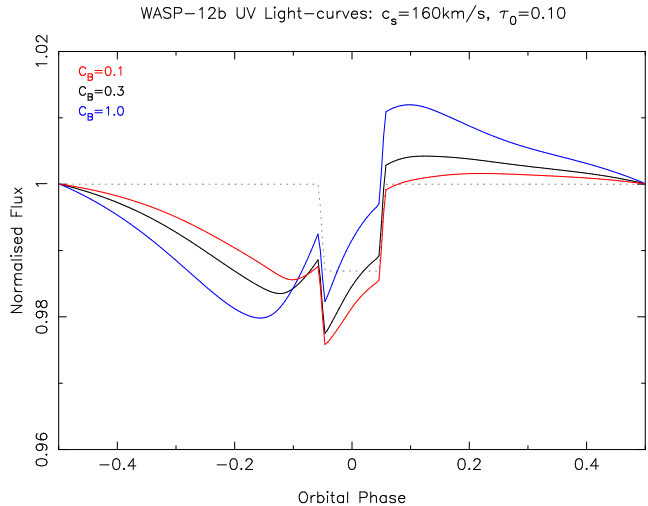}
       }
       \caption{As Fig.\,\ref{fig:light_curves_tau}, but for models with $c_{\mathrm s}$=160\kms, $\tau_0$=0.1, and various values of $C_{\mathrm B}$.  Stronger magnetic fields (i.e., larger magnetospheric cavities) increase the amplitude of the UV transit, and also result in a larger phase offset between the minima of the UV and optical transits.}
           \label{fig:light_curves_CB}
\end{figure}

\begin{figure}
\centering
       \resizebox{\hsize}{!}{
       \includegraphics[angle=270]{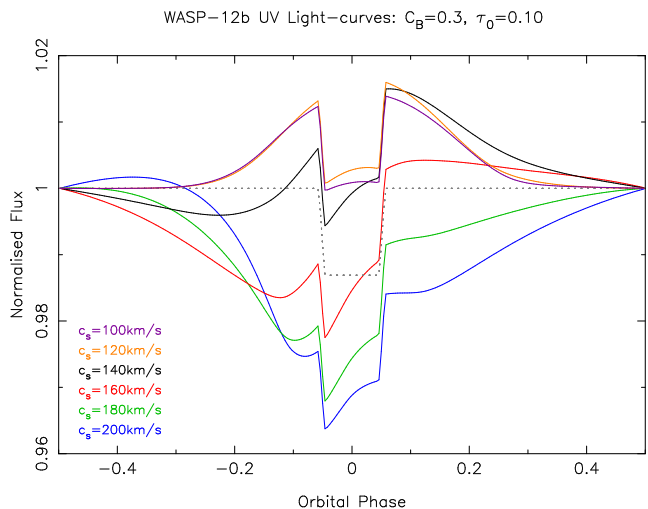}
       }
       \caption{As Figs.\,\ref{fig:light_curves_tau} \& \ref{fig:light_curves_CB}, but for models with $C_{\mathrm s}$=0.3, $\tau_0$=0.1, and various sound speeds.  For higher sound speeds the phase offset between UV and optical transits is smaller, as the bow shock is oriented towards the star (see Fig.\,\ref{fig:density_maps}).  In addition, the steep density profiles in the low-$c_{\mathrm s}$ wind solutions (see Fig\,.\ref{fig:parker_rho}) mean that for low sound speeds ($c_{\mathrm s}$$\lesssim$140\kms) most of the absorbing column lies interior to the planet's orbit.  In this case there is little (relative) absorption in the magnetosheath, and the radially-extended magnetospheric cavity instead results in a ``negative-depth'' transit.}
           \label{fig:light_curves_cs}
\end{figure}

As the density in the numerical simulations is scale-free, we define a normalised column density $N$ relative to the column density 180$^{\circ}$ out of phase with the planet:
\begin{equation}
N(\phi) = \frac{\Sigma(\phi)}{\Sigma(\phi_{\mathrm p}+\pi)} \, .
\end{equation}
As discussed above, we calculate the line-of-sight optical depth by assuming that the UV opacity of the wind gas is constant.  The UV intensity $I(\phi)$ is therefore given by
\begin{equation}
I(\phi) = I_{\mathrm p}(\phi)\, \mathrm e^{-\tau_0 N(\phi)} \, .
\end{equation}
Here the exponent is the line-of-sight optical depth $\tau$.  The optical depth parameter $\tau_0$ represents the mean optical depth of the gas in the wind (i.e., $\tau(\phi) = \tau_0 N(\phi)$), and $I_{\mathrm p}(\phi)$ is the light-curve of the planetary transit (i.e., the optical transit light-curve).  We neglect limb darkening and simply compute $I_{\mathrm p}(\phi)$ by assuming an opaque planet and a uniform brightness stellar disc, using Equation 1 of \citet{ma02} and the parameters of WASP-12b from \citet{hebb09}.  We plot all transit light-curves in units of normalised flux, relative to the intensity 180$^{\circ}$ out of phase with the planet [i.e., we plot $I(\phi)/I(\phi_{\mathrm p}+\pi)$].

Fig.\,\ref{fig:light_curves_tau} shows how the UV transit varies as a function of the optical depth parameter $\tau_0$ for a fiducial model (that with $c_{\mathrm s}$=160\kms\ and $C_{\mathrm B}$=0.3).  The magnetosheath provides significant absorption the stellar UV flux ahead of the optical transit, resulting in a broad ``dip'' in the UV light-curve ahead of the optical transit.  By contrast, the line-of-sight through magnetospheric cavity has a lower optical depth than the out-of-transit mean, so the normalised UV flux exceeds unity by a small amount ($<$1\%) in the post-transit region\footnote{This is in part due to our assumption of constant opacity. If the shocked gas has a higher mass absorption coefficient than the material in the wind, then this ``excess'' flux in the post-transit region will be negligible (see discussion in Section \ref{sec:previous}).}.  The minimum UV flux precedes the optical transit by 12--13\% in phase, which corresponds to a phase offset of $\simeq$3.5 hours for the 1.09d period of WASP-12b.  

Fig.\,\ref{fig:light_curves_CB} shows how the UV transit profiles change for different magnetospheric cavity sizes (i.e., different values of $C_{\mathrm B}$).  Increasing the magnetic field strength results in more pronounced features in the UV transit: the higher density in the bow shock causes increased absorption ahead of the optical transit, while the larger cavity results in a larger UV flux enhancement in the post-transit region (with values $C_{\mathrm B}$$\gtrsim$1.0 leading to flux enhancements of $>$1\%).  Larger cavities also subtend larger azimuthal angles, resulting in increased phase offsets between the UV absorption in the bow shock and the optical transit.  For $C_{\mathrm B}$=0.1 (i.e., a cavity size of $\simeq$6.0\Rp) the UV minimum is relatively weak, and occurs only marginally ahead of the optical transit ingress.  By contrast, for $C_{\mathrm B}$=1.0 (i.e., a cavity size of $\simeq$9.0\Rp) the minimum UV flux occurs at phase $-0.16$, which for WASP-12b corresponds to a phase offset of more than four hours. 

Similarly, Fig.\,\ref{fig:light_curves_cs} shows how the UV transit profiles depend on the sound speed in the wind, $c_{\mathrm s}$.  Here two different physical effects come into play.  First, as seen in Fig.\,\ref{fig:density_maps}, the geometry of the bow shock changes significantly with sound speed, with the magnetosphere oriented progressively more towards the star as the wind speed increases.  The azimuthal offset between the planet and peak density in the bow shock therefore decreases with increasing $c_{\mathrm s}$, and this is reflected in the transit profiles: the phase offset of the maximum absorption in the wind decreases from $\simeq$22\% ($\simeq$5.8 hours) for $c_{\mathrm s}$=140\kms\ to $\simeq$8\% ($\simeq$2.1 hours) for $c_{\mathrm s}$=200\kms.  Second, and more significantly, the radial density profile of the wind changes dramatically with $c_{\mathrm s}$ (see Fig.\,\ref{fig:parker_rho}), and is much steeper for low sound speeds.  For $c_{\mathrm s}$$\lesssim$140\kms\ most of the absorbing column is therefore interior to the planet's orbit, and the bow shock provides relatively little absorption.  In fact, for $c_{\mathrm s}$$\lesssim$140\kms\ the decrease in optical depth along lines-of-sight through to the cavity is larger than the increase through the magnetosheath, resulting in an increase in the relative UV flux and an unrealistic ``negative-depth'' transit.


\begin{figure}
\centering
       \resizebox{\hsize}{!}{
       \includegraphics[angle=270]{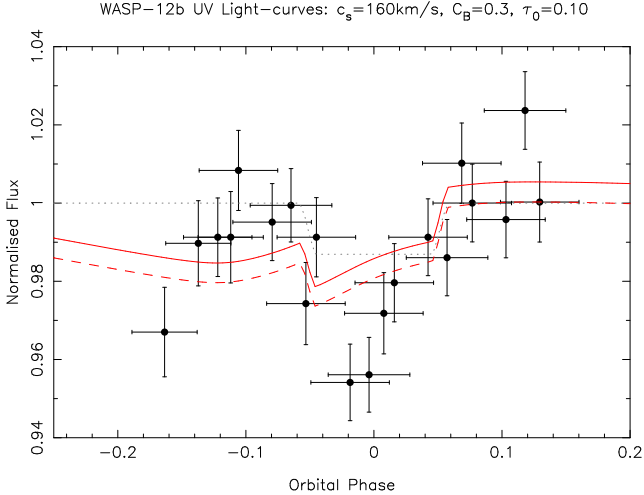}
       }
       \caption{Comparison between our fiducial model ($c_{\mathrm s}$=160\kms, $C_{\mathrm s}$=0.3, $\tau_0$=0.1) and recent near-UV observations of WASP-12b.  The black data points denote the fluxes reported in the {\it Hubble Space Telescope} study of \citet{nichols15}, normalised to the median flux (as in their Fig.3c): vertical error bars represent the Poisson errors on the observed fluxes, while the horizontal error bars denote the duration of the individual exposures.  The solid red line shows our fiducial model normalised to the median flux (i.e., in the same manner as the data from \citealt{nichols15}), while the dashed curve shows what we see if we instead normalise to the observed post-transit region ($\phi$=[0.08,0.10]), as in \citet{haswell12}. Our predicted light-curves are consistent with the observations, but the large error bars and limited phase coverage restrict our ability to place useful constraints on the model parameters.}
           \label{fig:light_curves_data}
\end{figure}

\begin{figure*}
\centering
       \resizebox{\hsize}{!}{
       \includegraphics[angle=270]{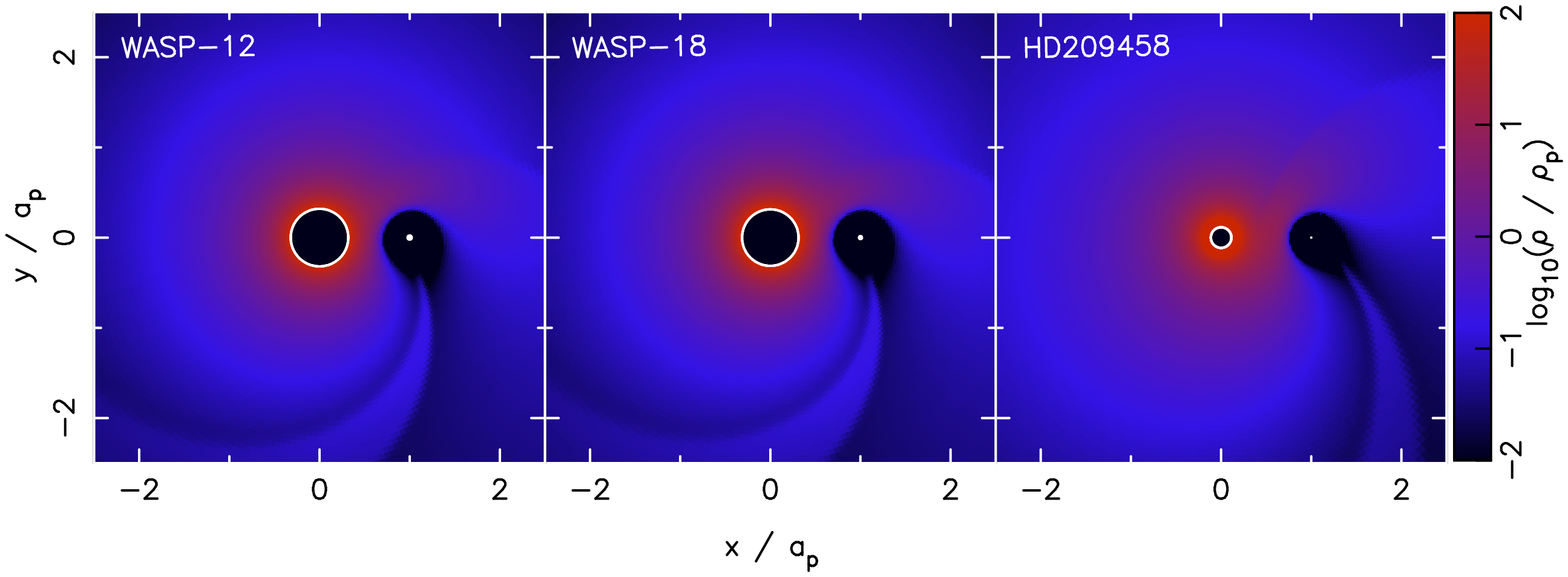}
       }
       \caption{Steady-state density structure in our simulations with $C_{\mathrm B} = 0.3$ and $c_{\mathrm s}$=160\kms, for WASP-12, WASP-18 \& HD209458. As in Fig.\ref{fig:density_maps}, the density colour-scale is normalised to the value at $a_{\mathrm p}$, and the planets and stars are plotted to scale.  The flow solution around the more massive WASP-18b has no significant differences from that around WASP-12b, showing that the planet's gravity is essentially negligible.  By contrast, the larger semi-major axis of HD209458b results in a different wind speed and ram pressure at the planet's orbit, leading to a larger magnetospheric cavity (relative to the star and planet) and the bow shock being angled significantly more towards the star.}
           \label{fig:density_maps_models}
\end{figure*}

Our results show that the magnetosheath has a clearly identifiable signature: a broad minimum in the UV light-curve which leads the optical transit by $\simeq$10--20\% in orbital phase.  Fig.\,\ref{fig:light_curves_data} shows a comparison between our fiducial model ($c_{\mathrm s}$=160\kms, $C_{\mathrm B}$=0.3, $\tau_0$=0.1) and the {\it HST} near-UV observations of WASP-12b by \citet{nichols15}.  The data are normalised to the median observed flux (as in Fig.\,3c of \citealt{nichols15}), as the limited phase coverage of the observations prevents us from normalising the observed light-curve in the same way as in Figs.\,\ref{fig:light_curves_tau}--\ref{fig:light_curves_cs}.  Our model light-curves are consistent with the observed near-UV transits of WASP-12b, and reproduce the main features of the observed light-curve: excess UV absorption during the optical transit, and a lower UV flux in the pre-transit region (compared to the post-transit region). However, the large error bars and partial phase coverage of the observations limit our ability to discriminate between different models.  Moreover, it is clear from from Figs.\,\ref{fig:light_curves_tau}--\ref{fig:light_curves_cs} that the predicted light-curves have significant degeneracies between the model parameters (wind sound speed/temperature, magnetic field strength and optical depth).  This suggests that transit light-curves obtained from broad-band UV observations alone cannot provide strong limits on these parameters. Wider phase coverage in particular is necessary if we are to break these degeneracies and establish a unique solution \citep[see also][]{bisikalo15}.


\section{Models of other systems}
To investigate how the observable signatures of the planet-wind interaction vary with model parameters, we also apply our model to two other transiting hot Jupiters: WASP-18b and HD209458b.  WASP-18b, a $\simeq$10\Mjup\ planet in a 0.94d period \citep{hellier09}, is one of the most massive known close-in exoplanets, and for our purposes represents a more massive analogue of WASP-12b.  By contrast HD209458b, the first transiting exoplanet to be discovered \citep{charbonneau00}, is a ~$\simeq$0.5\Mjup\ planet in a 3.5d period; here we use it as a test of moving the planet to a larger orbital separation.  Our adopted parameters for these systems are as follows.

\subsection*{WASP-18b}
We adopt the parameters from \citet{triaud10}: $M_*$=1.24\Msun, $a_{\mathrm p}$= 0.0202AU, $P$=0.94d (and therefore $u_{\mathrm p}$=233.4\kms), $R_*$=1.36\Rsun\ (=0.313\ap), $M_{\mathrm p}$=10.1\Mjup$=$$7.78\times10^{-3}M_*$ and $R_{\mathrm p}$=1.27\Rjup$=$0.0299\ap.  We maintain the same numerical resolution as before, with $N_{\phi}$=250; the computational grid spans $[r_{\mathrm {in}},r_{\mathrm {out}}]$=$[0.313a_{\mathrm p},10.0_{\mathrm p}]$, so $N_r$=140.

\subsection*{HD29458b}
We adopt the parameters from \citet{torres08}: $M_*$=1.12\Msun, $a_{\mathrm p}$=0.0471AU, $P$=3.52d (and therefore $u_{\mathrm p}$=145.2\kms), $R_*$=1.16\Rsun\ (=0.114\ap), $M_{\mathrm p}$=0.685\Mjup$=$$5.85\times10^{-4}M_*$ and $R_{\mathrm p}$=1.36\Rjup$=$0.0135\ap.  Here the grid spans $[r_{\mathrm {in}},r_{\mathrm {out}}]$=$[0.114a_{\mathrm p},10.0_{\mathrm p}]$, so with $N_{\phi}$=250 we require $N_r$=180.


\subsection{Results}
\begin{figure}
\centering
       \resizebox{\hsize}{!}{
       \includegraphics[angle=0]{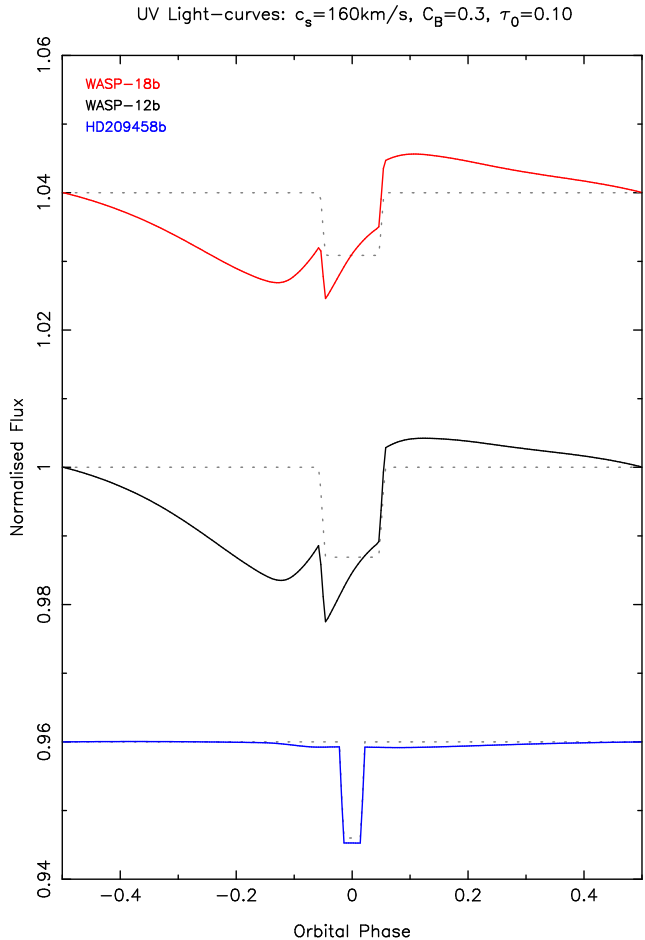}
       }
       \caption{Light curves for the three models shown in Fig.\,\ref{fig:density_maps_models} ($C_{\mathrm B} = 0.3$ and $c_{\mathrm s}$=160\kms); as before, the dotted lines denote the optical transits.  For clarity, the curves for WASP-18b (red) and HD209458b (blue) have been offset by $\pm$4\% in flux, respectively.  The light-curve for the more massive WASP-18b essentially identical to that of WASP-12b.  However, the larger orbital separation of HD209458b dramatically reduces the (relative) absorption in the bow shock, and the resulting UV light-curve is almost indistinguishable from the optical transit.}
           \label{fig:light_curves_models}
\end{figure}

In each case we run a new version of a fiducial model, with $c_{\mathrm s}$=160\kms\ and $C_{\mathrm B}$=0.3.  Fig.\,\ref{fig:density_maps_models} shows the steady-state flow solutions for both systems, as well as WASP-12b, while Fig.\,\ref{fig:light_curves_models} shows the corresponding UV light-curves.  In the case of WASP-18, the density structure and light-curve are both essentially indistinguishable from WASP-12.  This is despite the planet:star mass ratio being a factor of 7.8 larger for WASP-18, and demonstrates that the planet's gravity is negligible in determining the structure of the magnetosphere in hot Jupiters.  This is in stark contrast with the Roche lobe overflow model, which depends strongly on the planet's mass; we discuss the observational consequences of this result in Section \ref{sec:B_v_RLO} below.

HD209458, by contrast, shows pronounced differences from WASP-12 in both its flow solution and its transit light-curve.  The larger semi-major axis results in a higher wind speed at the planet's position, but also a much lower gas density.  Consequently the magnetosphere is much more extended (relative to both the star and planet) than for WASP-12b or WASP-18b.  The bow shock is also angled much more strongly towards the star (see Fig.\,\ref{fig:density_maps_models}), primarily as a result of the planet's lower orbital velocity.  However, the most significant effect of the larger orbital separation in HD209458 is that line-of-sight column density through the wind is dominated by gas well inside the planet's orbit.  The magnetosheath and cavity contribute only a very small fraction of the total column, and consequently are almost undetectable: the UV light-curve is essentially indistinguishable from the optical transit (see Fig.\,\ref{fig:light_curves_models}).  This is consistent with existing UV observations of HD209458 \citep{v-m13}, and suggests that absorption of stellar UV in the magnetosheaths of hot Jupiters will be undetectable for planets with periods $\gtrsim$1.5d\footnote{This is further supported by recent observations of WASP-13b, which has a period of 4.35d and shows no evidence for circumplanetary absorption \citep{fossati16}.}.

\section{Near-UV absorption in the stellar wind}\label{sec:mocassin}
We now address the thorny question of the origin of this absorption, and the magnitude of the opacity in the magnetosheath. Thus far we have assumed a constant opacity $\kappa$, and parametrized our light-curves in terms of the mean (azimuthally-averaged) optical depth $\tau_0$. In reality the continuum near-UV (2000--3000\AA) opacity of $\sim$$10^6$K gas is due to a complex blend of thousands of lines of metal ions \citep[e.g.,][]{opal96}, and each individual line can be very sensitive to both density and temperature. It is relatively straightforward to compute the opacity for individual spectral lines \citep[e.g.,][]{lai10}, but calculating the ``continuum'' opacity from the blanketed, blended line forest is extremely challenging: even state-of-the-art radiative transfer models are still subject to significant uncertainties in this region. However, it is striking that existing UV observations of WASP-12b show significant absorption of the stellar continuum across the entire near-UV waveband, with roughly constant levels of (relative) absorption in a number of different UV spectral regions \citep{fossati10,haswell12,nichols15}.  This suggests that the integrated ``continuum'' opacity (away from strong resonance lines), which is due to many thousands of individual spectral lines, is not a strong function of wavelength.  Our assumption of constant opacity is therefore a reasonable first-order approximation (and so the shapes of our predicted light-curves are robust), but further calculations are needed to quantify the magnitude of the near-UV absorption.

The most useful constraints on the absorbing column come from the the Mg\,{\sc ii} resonance lines at 2800\AA: these lines are among the strongest in the near-UV spectral band, and are therefore the most sensitive to small absorbing columns. For optical depth unity in the line centre, \citet{lai10} estimated a critical column density of $1.3\times10^{13}$cm$^{-2}$, which requires a Mg$^+$ number density in the wind of $n_{\mathrm {MgII}}\simeq400$cm$^{-3}$. More recent observations of WASP-12 show that the line core is in fact extremely optically thick during the transit, and the true absorbing column may be as high as $10^{17}$cm$^{-2}$ \citep{haswell12}. We therefore treat $n_{\mathrm {MgII}}=400$cm$^{-3}$ as a lower limit: if the stellar wind and magnetosheath is is responsible for the observed absorption, the local density of Mg$^+$ must exceed this value.

Estimating the density of Mg$^+$ in the wind is essentially a problem of ionization balance, so to investigate this issue in more detail we use the photoionization code {\sc mocassin} \citep{ercolano03,ercolano05,ercolano08}. We computed a series of models in which an isothermal, constant-density column of gas (i.e., a 1-D ``slab'') with solar abundances is irradiated by a standard (solar) coronal spectrum. The grid of models spans the ranges $n = 10^3$\ldots$10^7$cm$^{-3}$ and $\log_{10}(T/\mathrm K) = 4.0$\ldots6.5, and the slabs are all $10^{12}$cm thick\footnote{For a wind rate $\dot{M}_{\mathrm w}$=$10^{-15}$\Msunyr\ and $c_{\mathrm s}$=160\kms\ (i.e., our fiducial model), the number density in the (unperturbed) wind at the planet's orbital radius is $n(a_{\mathrm p}) = 2.7\times10^3$cm$^{-3}$.}. In practice the gas has low optical depth, so the ionization balance is primarily determined by the local temperature and density, and varies negligibly across the computational grid. For convenience we use the results calculated at the centre of the grid.

\begin{figure}
\centering
       \resizebox{\hsize}{!}{
       \includegraphics[angle=270]{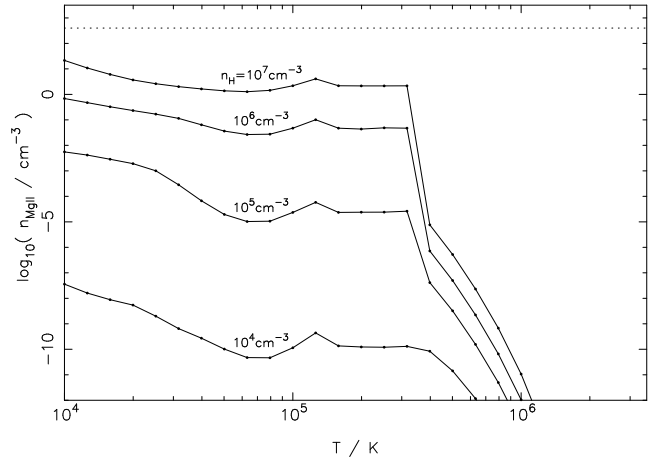}
       }
       \caption{Number density of Mg\,{\sc ii} as a function of density and temperature, calculated using {\sc mocassin} (the small dots represent the individual slab models). The precipitous drop in $n_{\mathrm {MgII}}$ for $T \gtrsim 3\times10^5$K is due to the thermal ionization of Mg$^+$. The dashed line shows the critical value of $n_{\mathrm {MgII}}=400$cm$^{-3}$, required for optical depth unity in the Mg\,{\sc ii} resonance lines \citep{lai10}. Even in the most optimistic case, the Mg\,{\sc ii} density in the stellar wind remains at least an order of magnitude below this threshold.}
           \label{fig:mocassin}
\end{figure}

Fig.\,\ref{fig:mocassin} shows how the Mg\,{\sc ii} number density varies as a function of density and temperature.  For a given gas density we see that $n_{\mathrm {MgII}}$ varies only weakly with temperature between $T=10^4$--$3\times10^5$K, but declines precipitously (by many orders of magnitude) for $T>3\times10^5$K. This is due to the thermal ionization of Mg$^+$, and is largely independent of the gas density. The $\sim$$10^6$K gas in the wind therefore provides essentially zero absorption in the Mg\,{\sc ii} resonance lines. However, we also note that even in the most optimistic case of $n = 10^7$cm$^{-3}$ (which corresponds to a wind rate $\dot{M}_{\mathrm w}\simeq10^{-12}$\Msunyr) and $T=10^4$K, the Mg\,{\sc ii} density is still an order of magnitude below the required value. We therefore conclude that achieving the required levels of absorption in the magnetosheath requires both a very high mass-loss rate in the stellar wind, and very rapid cooling of the shocked gas.

The radiative cooling time-scale in the shock is given by
\begin{equation}\label{eq:t_cool}
t_{\mathrm {cool}} = \frac{n k_{\mathrm B} T}{n_{\mathrm e} n_{\mathrm i} \Lambda} \, ,
\end{equation}
where $n_{\mathrm e}$ and $n_{\mathrm i}$ are the electron and ion number densities, $k_{\mathrm B}$ is Boltzmann's constant, and $\Lambda(n,T)$ is the (density-weighted) cooling function. For ionized gas we have $n \simeq n_{\mathrm e} \simeq n_{\mathrm i}$, so Equation \ref{eq:t_cool} simplifies to
\begin{equation}
t_{\mathrm {cool}} \simeq \frac{k_{\mathrm B} T}{n \Lambda} \, .
\end{equation}
For $T \sim 10^5$--$10^{6.5}$K and solar abundances, gas cooling is dominated by metal line emission and the cooling function $\Lambda$ varies only weakly with density \citep{sd93}. By adopting the cooling function from \citet{sd93}, we see that $\Lambda \sim 10^{-22}$erg\,s$^{-1}$\,cm$^{-3}$ for $T \sim 10^6$K, so the radiative cooling time-scale in the magnetosheath is of order
\begin{equation}
t_{\mathrm {cool}} \sim 10^8\,\mathrm s \, \left(\frac{n}{10^4\,\mathrm {cm}^{-3}}\right)^{-1} \, .
\end{equation}
The typical flow time-scale $a_{\mathrm p}/u_{\mathrm w} \sim 10^5$s\,$\ll t_{\mathrm {cool}}$, so in our fiducial model radiative cooling is negligible: the cooling time-scale exceeds the flow time-scale by 3 orders of magnitude. Radiative cooling is only significant for gas densities $\gtrsim 10^7$cm$^{-3}$ which, as noted above, correspond to extremely high rates of mass-loss in the stellar wind ($\dot{M}_{\mathrm w}\gtrsim10^{-12}$\Msunyr). If the stellar wind and/or planetary magnetosheath is to account for the observed absorption in the Mg\,{\sc ii} resonance lines seen in WASP-12, it must therefore cool non-radiatively (perhaps through loss of energy to the magnetic field). However, in the absence of any such non-radiative cooling, we conclude that it is unlikely that the observed absorption occurs in the magnetospheric bow shock.


\section{Discussion}\label{sec:dis}

\subsection{Limitations}\label{sec:limitations}
The hydrodynamic models presented here are somewhat simplified, and neglect several potentially important issues.  First, we neglect both motion and rotation of the central star.  Neglecting the motion of the star by fixing the planetary orbit is a good approximation, and we expect significant ``reflex'' stellar motion only in the case of WASP-18 (where $M_{\mathrm p}/M_*$=7.78$\times10^{-3}$).  However, even in this case the star's orbital radius (around the system barycentre) is at least an order of magnitude smaller than the magnetospheric radius, so this is unlikely to alter our results significantly.  Similarly, although \citet{vidotto10} argue that rotation of the stellar wind may be important, we find that even very fast stellar rotation has a negligible effect on our results.  For WASP-12 a stellar rotation period of 3d corresponds to a rotational velocity of 26\kms, and if angular momentum is conserved in the wind we expect the azimuthal component of the wind velocity at \ap\ to be $<$10\kms, much smaller than the planet's orbital velocity ($u_{\mathrm p}$=228\kms).  Even in the extreme limit of a magnetised wind which co-rotates with the star, the azimuthal component of the wind velocity at \ap\ is $\simeq$80\kms, and the effective Mach number in the planet's frame is reduced by only 10--20\%.  Rapid stellar rotation therefore represents only a small perturbation to the wind structure, and test calculations indicate that it has no measurable effect on our predicted light-curves.

The major simplification in our hydrodynamic calculations is the treatment of magnetic fields.  As discussed in Section \ref{sec:planet}, we do not run full MHD simulations, and instead model the planetary magnetic field as a spherically-symmetric acceleration that scales $\propto |\mathbf r - \mathbf r_{\mathrm p}|^{-7}$.  This approximation is strictly valid only for a dipole magnetic field which is aligned with the planet's orbit.  However, at the radii of interest ($|\mathbf r - \mathbf r_{\mathrm p}|\simeq5$--25\Rp) the dipole component of the field is dominant, and in this regime even a strongly inclined field will result in only a moderately elliptical magnetosphere.  Moreover, the integrated column density (and therefore the predicted light-curve) is not particularly sensitive to the precise shape of the magnetospheric shock and cavity, so a more realistic treatment of the planet's magnetic field is unlikely to change our results significantly.

A potentially more important simplification is neglecting the stellar magnetic field.  In the general case, the magnetospheric radius $r_{\mathrm m}$ (strictly, the radius of the head of the magnetopause) is given by \citep[e.g.,][]{lai10}
\begin{equation}
\rho u^2 + \rho c_{\mathrm s}^2 + \frac{B_{\mathrm w}^2}{8\pi} = P_{\mathrm p} + \frac{B_0^2}{8\pi}\left(\frac{R_{\mathrm p}}{r_{\mathrm m}}\right)^6 \, .
\end{equation}
The first and second terms on the left-hand-side, respectively, are the ram and thermal pressures of the wind gas, while the final term on the right-hand-side is the pressure from the planetary magnetic field; all of these are included explicitly in our simulations.  $P_{\mathrm p}$ is the gas pressure inside the magnetosphere; this is neglected in our models, but at the radii of interest ($>$5--10\Rp) it is unlikely to be significant unless the planet is losing significant mass via Roche lobe overflow (see Section \ref{sec:B_v_RLO} below), or the magnetosphere is filled with plasma from another source (such as moons or rings).  The third term on the left-hand side is the magnetic pressure in the wind, due to stellar magnetic field lines threading the wind ($B_{\mathrm w}$ is the magnetic field strength in the wind).  The magnitude of the stellar magnetic field in hot Jupiter host stars is not well constrained, but comparison to the Solar wind suggests that the magnetic pressure in the wind could be comparable to the gas pressure \citep[see discussion in][]{lai10}.  We do not explicitly include this term in our models, but our adopted parametrization (using accelerations rather than pressures) means that only the gradient of the pressure $\partial/\partial r(B_{\mathrm w}^2)$ is important;  a constant magnetic pressure $B_{\mathrm w}^2/8\pi$ simply represents an offset in the scaling relation between our input parameter $C_{\mathrm B}$ and the planetary magnetic field $B_0$ (Equation \ref{eq:B_CB}).  The magnetic pressure in the wind is unlikely to vary dramatically across the magnetosphere, so neglecting this term has only a small effect on the magnetospheric geometries in our simulations.

Our estimates of the planetary field strength $B_0$, however, do depend on the absolute value of $B_{\mathrm w}$.  If magnetic pressure in the wind is significant then Equation (\ref{eq:B_CB}) represents a lower limit on $B_0$ for a given input parameter $C_{\mathrm B}$ (or, alternatively, a given magnetospheric radius).  With a fiducial stellar wind rate of $\dot{M}_{\mathrm w} = 10^{-15}$\Msunyr\ and our adopted values of $C_{\mathrm B}$, Equation (\ref{eq:B_CB}) gives surface planetary field strengths spanning the range $B_0$$\simeq$2--25G (corresponding to pressures of 0.2--2$\times10^{-5}$g\,cm$^{-1}$\,s$^{-2}$); the true values could be higher or lower, depending on the contributions from magnetic pressure in the wind and gas pressure in the cavity.  However, unless one of these terms (which we have neglected) dominates the dynamics, which seems unlikely, our simulations should give qualitatively correct structures.  Moreover, our predicted light-curves depend primarily on the size of the magnetospheric cavity and the density in the bow-shock, both of which are largely insensitive to the absolute value of $B_{\mathrm w}$ (unless the field has a very unusual geometry, such as a steep magnetic pressure gradient across the magnetosphere).  Finally, we note that the flow in the low-density wake of the magnetosphere is likely to form an extended magnetotail containing open field lines, which will periodically reconfigure in explosive reconnection events. Moreover, evaporation of material from the planet dominates the structure of the tail even for relatively low outflow rates \citep{matsakos15}. Consequently we do not consider our (steady-state) simulations to be accurate here. However, in most cases this region does not contribute significantly to the integrated column density $N$, and consequently our predicted light-curves do not depend strongly on the structure of the wake.

The other obvious simplification in our approach is that our simulations are 2-D, while the transit of a magnetospheric bow shock is intrinsically a 3-D phenomenon.  Here our motivation is simple: 3-D hydrodynamic simulations remain computationally expensive, and when combined with the (much) larger parameter space for 3-D models this makes a large ensemble of simulations unfeasible.  The accuracy of the 2-D approximation is greatly increased in this case, however, because (somewhat fortuitously) the observational constraints on the WASP-12b system require the magnetospheric radius to be comparable to the stellar radius \citep[e.g.,][]{lai10,vidotto10}.  The UV transit therefore involves two objects with comparable angular sizes, and this is much less sensitive to, for example, inclination, than a transit of objects of very different sizes.  Additional consideration of this issue is obviously required, but test calculations suggest that moving to 3-D is unlikely to change our results dramatically.

Finally, the calculations of the near-UV opacity in Section \ref{sec:mocassin} are obviously rather simplified, but the result is so clear-cut ($n_{\mathrm {MgII}}$ is $\sim$$10^{15}$ below the critical value) that these simplifications are negligible.  At $T\sim10^6$K the coronal gas is simply too highly ionized for any Mg$^+$ to be present; the only way to account for the observed absorption is to assume that the shocked gas is cold ($T\sim10^4$K). In this case our simulated light-curves will no longer be accurate, as our assumption of constant opacity will over-estimate the absorption in the un-shocked gas. If the material in the wind has significantly lower opacity than the shocked gas in the magnetosheath we would predict negligible absorption out of transit (i.e., from orbital phase $\simeq 0.2$ to $-0.3$ in Figs.\,\ref{fig:light_curves_tau}--\ref{fig:light_curves_data}), and the resulting changes in the normalisation would also eliminate the ``excess'' emission from the post-transit region. Further support for the presence of cold absorbing gas comes from the recent observations of HD189733 by \citet{cauley15}, who detected several Balmer absorption lines (from atomic hydrodgen) during transit. \citet{lai10} considered $T\sim10^4$K to be ``the most optimistic scenario'' but did not discuss the cooling mechanism(s), and we have shown that radiative cooling is inefficient at the densities and temperatures expected in the magnetosheath. Moreover, even if the gas can cool, we must still invoke very high stellar wind rates ($\dot{M}_{\mathrm w}\gtrsim10^{-12}$\Msunyr) if the absorbing column is to be large enough to reproduce the observed Mg\,{\sc ii} resonance lines at 2800\AA. As a result, although our simulated light-curves show qualitatively good agreement with the observed near-UV transit data, we do not consider absorption in the magnetospheric bow shock to be a likely explanation.


\subsection{Comparison to previous studies}\label{sec:previous}
Our models are among the first to apply self-consistent hydrodynamics to this problem (absorption in  the magnetosheaths of hot Jupiters), and our flow solutions and light-curves differ significantly from previous studies \citep{vidotto10,vidotto11,llama11,llama13}.  These studies did not calculate the shock structure explicitly.  Instead they assumed that the shock was in the hyper-sonic limit ($\mathcal M \gg 1$) and applied the Rankine--Hugoniot jump conditions at the magnetospheric radius. This assumption of a strong shock in turn implies that the shocked region is narrow: for WASP-12b, \citet{llama11} consider normalised shock widths ($\Delta r_{\mathrm m}/r_{\mathrm m}$) ranging from 0.01--0.06.  However, our hydrodynamic calculations show that in this case the strong shock assumption is not valid.  When we compute the wind dynamics self-consistently we find that the shocks are only marginally supersonic ($\mathcal M\simeq1.6$--1.8; see Section \ref{sec:sims}), with the hyper-sonic limit being reached only for unrealistically low sound speeds ($c_{\mathrm s} \lesssim 50$\kms).  The resulting bow shocks are invariably weak, with widths 10-100 times larger ($\Delta r_{\mathrm m}/r_{\mathrm m}\simeq1.5$--2) than assumed by \citet{vidotto10,vidotto11} and \citet{llama11,llama13}.  The weak shocks also have lower density contrasts than assumed in the hyper-sonic limit, and the large shock widths mean that the magnetospheric structures typically extend around $\gtrsim 30\%$ of the planet's orbit.  This in turn leads to much broader features in our predicted light-curves: for $\tau_0 \sim 0.1$, the minimum of the UV light-curve typically leads the optical transit by $>$10\% in phase, and significant absorption is seen over a large fraction of the planet's orbit.  The extended magnetosheaths in our simulations differ substantially from the magnetospheric structures assumed in previous studies, and the differences between our results and those of \citet{vidotto10,vidotto11} and \citet{llama11,llama13} are primarily due to these authors' inaccurate assumption of strong shocks (with $\mathcal M \gg 1$).

The other main difference between our models and those of \citet{vidotto10,vidotto11} and \citet{llama11,llama13} lies in our treatment of the absorption. Our weak magnetospheric shocks are comparable in size to the star, with large covering fractions (of order unity) and only moderate (per cent level) increases in the absorbing column along the line-of-sight through the magnetosheath.  By contrast, the physically narrow shocks (inappropriately) assumed by \citet{vidotto10,vidotto11} have small covering fractions, and must therefore have much higher optical depths if they are to absorb stellar UV flux at the per cent level. This can only be achieved by assuming that the shocked gas cools to $\sim10^4$K. The original study of \citet{lai10} considered the possibility of cooling to $\sim10^4$K as the ``most optimistic'' case, but they and subsequent studies \citep[e.g.,][]{vidotto10,vidotto11,llama11,llama13} did not discuss the cooling mechanism(s) in detail. We have shown that radiative cooling is inefficient (as the radiative cooling time-scale is invariably much longer than the flow time-scale), so it is not at all clear that the assumption of a $\sim10^4$K shock is justified. There are further minor differences in how we generate light-curves from our simulations \citep[][for example, use a Monte Carlo radiative transfer code to generate synthetic light-curves; we also neglect limb darkening]{llama11}, but for a given optical depth our treatment of absorption in the shocked gas is essentially the same as that of \citet{llama13}, so the simplifications we make here are unlikely to have a significant effect on our results. In the absence of a physical mechanism to cool the shocked gas rapidly (from $\sim10^6$K to $\sim10^4$K, on a time-scale $\lesssim 10^5$s), we conclude that the magnetospheric bow shock cannot account for the excess absorption seen in near-UV observations of WASP-12. 

The most popular alternative explanation for the broad UV transit of WASP-12b is Roche lobe overflow.  This idea was first suggested by \citet{lai10}, and has subsequently been explored in more detail by \citet{bisikalo13} and \citet{tripathi15}.  The main obstacle here is that the radius of WASP-12b (1.79$\pm$0.09\Rjup, derived from the optical transit by \citealt{hebb09}) is smaller than the Roche lobe radius\footnote{Using the \citet{eggleton83} formula, the radius of a sphere of equal volume to the Roche lobe of WASP-12b is $R_1$=2.37\Rjup.} by a factor of $\simeq$1.3.  Roche lobe overflow therefore requires the planet to have an inflated atmosphere, extending $>$30\% above the planet's optical radius.  This is possible if the planet is strongly irradiated by, for example, high-energy photons from the star \citep*[e.g.,][]{lopez12,oj12,tripathi15}, and naturally produces cold gas to absorb the stellar UV flux, but it is not clear whether WASP-12 is sufficiently luminous at these high energies to drive significant mass-loss.  (Note that the models of \citealt{bisikalo13} assume the presence of a $10^4$K atmosphere as a boundary condition.)  To date, detailed light-curves for the transits of Roche-lobe-overflowing hot Jupiters have not been computed, and current data do not distinguish between this model and absorption in a magnetospheric bow shock.  However, our models suggest that observations of more massive planets, such as WASP-18b, may provide a straightforward test and distinguish between these two scenarios cleanly.  We discuss this idea in more detail in Section \ref{sec:B_v_RLO} below.

In reality it is likely that both of these processes operate simultaneously. \citet{matsakos15} recently presented a suite of simulations of star-planet interactions which incorporate stellar winds, planetary evaporation, and magnetic fields. They classify the resulting flows into four types, depending on the relative importance of the planetary magnetic field, planetary outflow and stellar gravity. If we adopt the same classification scheme we see that our simulations all fall into the ``Type I'' regime, where the planetary outflow is weak. We see broadly similar flow structures to the MHD simulations of \citet{matsakos15}, despite the fact that we neglect planetary outflow completely, but it is notable that in their calculations the structure of the tail/wake is dominated by material outflowing from the planet (even for low outflow rates). \citet{matsakos15} find that the planetary outflow is typically ionized, but may well remain cold enough to provide significant absorption in the Mg\,{\sc ii} resonance lines. Future calculations of magnetospheric absorption should therefore consider the dynamics of the planetary outflow in more detail.

Finally, we note that the broad absorption features in our light-curves have important implications for the interpretation of UV observations.  If the magnetosheath can cool sufficiently to provide substantial opacity in the near-UV, then the broad, extended shock results in significant absorption (comparable to the depth of the optical transit) over a large fraction of the planet's orbit. We also predict a significant UV flux asymmetry either side of the optical transit.  Existing UV observations have had limited phase coverage (e.g., the data in \citet{haswell12} and \citet{nichols15} cover the phase range $\simeq$$-$0.16 to $+$0.12), and the extreme ``out-of-transit'' points are typically used for normalisation.  However, our simulations show that this is insufficient to measure a true out-of-transit UV flux, and suggest that ``de-trending'' across this limited phase range may introduce significant artefacts to the observed light-curves \citep[see also the discussion in][]{nichols15}. Wider phase coverage, preferably extending to at least $\pm$0.25 in orbital phase, is therefore highly desirable if future UV observations are to provide further insight into the environments of planets such as WASP-12b.


\subsection{WASP-12 vs.~WASP-18: a critical test of UV absorption theories}\label{sec:B_v_RLO}
At present there are two competing explanations in the literature for the excess UV absorption seen in WASP-12: Roche lobe overflow \citep{lai10,bisikalo13} and a magnetospheric bow shock \citep{lai10,vidotto10}. Our consideration of the opacity (see Section \ref{sec:mocassin}) suggests that cold ($T\sim10^4$K) Roche lobe overflow is a more plausible near-UV absorber than the hot ($T\sim10^6$K) magnetosheath, but current data do not allow us to distinguish between these models. However, even in the absence of a first-principles calculation of the gas opacity, our simulations point to a critical test that will provide a clear answer to this question.  WASP-18 has a comparable stellar mass to WASP-12, and their planets have similar radii and orbital separations.  In magnetospheric shock models the planet's gravity is negligible, and the shape of the UV light-curve depends primarily on the sound speed in the wind and the spatial extent of the magnetosphere.  Our predicted light-curves for WASP-18 are therefore indistinguishable from those for WASP-12 (see Fig.\,\ref{fig:light_curves_models}).  However, as WASP-18b is 7.2 times more massive than WASP-12b, it has a much larger Roche lobe ($R_1$=3.2\Rp, compared to 1.3\Rp\ for WASP-12b).  In the Roche overflow model the mass-flux through the L1 point decreases exponentially with increasing Roche lobe radius, and significant mass-loss is essentially impossible for a planet as massive as WASP-18b.  

Taken together, these results imply that UV observations of WASP-18 represent a straightforward test for these models.  A UV transit of WASP-18b which looks similar to that of WASP-12b (i.e., showing excess UV absorption) would argue strongly against the Roche lobe overflow model, as Roche lobe overflow cannot provide significant mass-loss from WASP-18b.  By contrast, the absence of excess UV absorption in WASP-18b would suggest that magnetospheric absorption is not significant.  Both processes may occur simultaneously in the WASP-12 system, and independent measurements of the planetary magnetic fields are required for a definitive test the magnetospheric bow shock hypothesis.  However, a clear detection of excess UV absorption in WASP-18 would essentially rule out the Roche lobe overflow hypothesis, and point strongly towards magnetospheric absorption as the most likely explanation.


\section{Summary}\label{sec:concs}
We have presented hydrodynamic simulations of stellar wind--magnetosphere interactions in hot Jupiters such as WASP-12b.  We work within an existing theoretical picture \citep[e.g.,][]{lai10,vidotto10,llama11}, but use numerical hydrodynamics to compute the wind and shock structure self-consistently.  We find that the structure of the magnetospheric bow shock differs substantially from that assumed by \citet{vidotto10,vidotto11} and \citet{llama11,llama13}.  For fiducial stellar wind rates we find that a planetary magnetic field of order a few G results in an extended magnetospheric cavity around the planet, typically 6--9\Rp\ in radius.  In the frame of the planet the stellar wind is always supersonic, leading to a bow shock ahead of the magnetosphere, but the Mach number is modest ($\mathcal M \simeq 1.6$--1.8) and consequently the shock is weak and broad.  The planet's magnetic field therefore creates a large perturbation to the wind, which typically extends around $\gtrsim30$\% of the planet's orbit, and the increased gas density in the shock can lead to in increased UV absorption.

We have used our simulations to generate synthetic transit light-curves (for a parametrized optical depth), and find that the weak bow shock has a characteristic signature in the UV light-curve: broad excess absorption which leads the optical transit by 10--20\% in orbital phase.  However, we require a near-UV optical depth $\tau \sim 0.1$ to explain the observed absorption in WASP-12b, and it is not clear how this can be achieved.  The $\sim10^6$K stellar wind has insufficient opacity (by many orders of magnitude) to account for the absorption seen in the Mg\,{\sc ii} resonance lines, and we find that this model can only produce significant near-UV absorption if the gas cools to $\sim10^4$K. Radiative cooling is inefficient, however, so we conclude that the magnetospheric bow shock is unlikely to be the origin of the observed near-UV absorption. 

We have also applied our model to two other hot Jupiters (WASP-18b and HD209458b). Regardless of the source of the opacity, we find that UV observations of WASP-18b (which is much more massive than WASP-12b) should provide a straightforward test to distinguish between Roche lobe overflow and a magnetospheric bow shock. Finally, our results also suggest that wider phase coverage is highly desirable in future such observations. 

\section*{Acknowledgements}
We thank Walter Dehnen for useful discussions, and an anonymous referee for a thoughtful review. RDA and JDN both acknowledge support from Science \& Technology Facilities Council (STFC) Advanced Fellowships (ST/G00711X/1 \& ST/1004084/1, respectively). RDA also acknowledges support from The Leverhulme Trust though a Philip Leverhulme Prize. Astrophysical research at the University of Leicester is supported by an STFC Consolidated Grant (ST/K001000/1).  This research used the ALICE High Performance Computing Facility at the University of Leicester. Some resources on ALICE form part of the DiRAC Facility jointly funded by STFC and the Large Facilities Capital Fund of BIS.  This work also used the DiRAC {\it Complexity} system, operated by the University of Leicester IT Services, which forms part of the STFC DiRAC HPC Facility ({\tt http://www.dirac.ac.uk}). This equipment is funded by BIS National E-Infrastructure capital grant ST/K000373/1 and  STFC DiRAC Operations grant ST/K0003259/1. DiRAC is part of the UK National E-Infrastructure.


\begin{thebibliography}{}
\makeatletter
\relax
\def\mn@urlcharsother{\let\do\@makeother \do\$\do\&\do\#\do\^\do\_\do\%\do\~}
\def\mn@doi{\begingroup\mn@urlcharsother \@ifnextchar [ {\mn@doi@}
  {\mn@doi@[]}}
\def\mn@doi@[#1]#2{\def\@tempa{#1}\ifx\@tempa\@empty \href
  {http://dx.doi.org/#2} {doi:#2}\else \href {http://dx.doi.org/#2} {#1}\fi
  \endgroup}
\def\mn@eprint#1#2{\mn@eprint@#1:#2::\@nil}
\def\mn@eprint@arXiv#1{\href {http://arxiv.org/abs/#1} {{\tt arXiv:#1}}}
\def\mn@eprint@dblp#1{\href {http://dblp.uni-trier.de/rec/bibtex/#1.xml}
  {dblp:#1}}
\def\mn@eprint@#1:#2:#3:#4\@nil{\def\@tempa {#1}\def\@tempb {#2}\def\@tempc
  {#3}\ifx \@tempc \@empty \let \@tempc \@tempb \let \@tempb \@tempa \fi \ifx
  \@tempb \@empty \def\@tempb {arXiv}\fi \@ifundefined
  {mn@eprint@\@tempb}{\@tempb:\@tempc}{\expandafter \expandafter \csname
  mn@eprint@\@tempb\endcsname \expandafter{\@tempc}}}

\bibitem[\protect\citeauthoryear{{Bisikalo}, {Kaygorodov}, {Ionov},
  {Shematovich}, {Lammer}  \& {Fossati}}{{Bisikalo} et~al.}{2013}]{bisikalo13}
{Bisikalo} D.,  {Kaygorodov} P.,  {Ionov} D.,  {Shematovich} V.,  {Lammer} H.,
   {Fossati} L.,  2013, \mn@doi [\apj] {10.1088/0004-637X/764/1/19}, \href
  {http://adsabs.harvard.edu/abs/2013ApJ...764...19B} {764, 19}

\bibitem[\protect\citeauthoryear{{Bisikalo}, {Kaigorodov}  \&
  {Konstantinova}}{{Bisikalo} et~al.}{2015}]{bisikalo15}
{Bisikalo} D.~V.,  {Kaigorodov} P.~V.,   {Konstantinova} N.~I.,  2015, \mn@doi
  [Astronomy Reports] {10.1134/S1063772915090012}, \href
  {http://adsabs.harvard.edu/abs/2015ARep...59..829B} {59, 829}

\bibitem[\protect\citeauthoryear{{Bodenheimer}, {Lin}  \&
  {Mardling}}{{Bodenheimer} et~al.}{2001}]{bodenheimer01}
{Bodenheimer} P.,  {Lin} D.~N.~C.,   {Mardling} R.~A.,  2001, \mn@doi [\apj]
  {10.1086/318667}, \href {http://adsabs.harvard.edu/abs/2001ApJ...548..466B}
  {548, 466}

\bibitem[\protect\citeauthoryear{{Cauley}, {Redfield}, {Jensen}, {Barman},
  {Endl}  \& {Cochran}}{{Cauley} et~al.}{2015}]{cauley15}
{Cauley} P.~W.,  {Redfield} S.,  {Jensen} A.~G.,  {Barman} T.,  {Endl} M.,
  {Cochran} W.~D.,  2015, \mn@doi [\apj] {10.1088/0004-637X/810/1/13}, \href
  {http://adsabs.harvard.edu/abs/2015ApJ...810...13C} {810, 13}

\bibitem[\protect\citeauthoryear{{Charbonneau}, {Brown}, {Latham}  \&
  {Mayor}}{{Charbonneau} et~al.}{2000}]{charbonneau00}
{Charbonneau} D.,  {Brown} T.~M.,  {Latham} D.~W.,   {Mayor} M.,  2000, \mn@doi
  [\apjl] {10.1086/312457}, \href
  {http://adsabs.harvard.edu/abs/2000ApJ...529L..45C} {529, L45}

\bibitem[\protect\citeauthoryear{{Cranmer}}{{Cranmer}}{2004}]{cranmer04}
{Cranmer} S.~R.,  2004, \mn@doi [American Journal of Physics]
  {10.1119/1.1775242}, \href
  {http://adsabs.harvard.edu/abs/2004AmJPh..72.1397C} {72, 1397}

\bibitem[\protect\citeauthoryear{{Eggleton}}{{Eggleton}}{1983}]{eggleton83}
{Eggleton} P.~P.,  1983, \mn@doi [\apj] {10.1086/160960}, \href
  {http://adsabs.harvard.edu/abs/1983ApJ...268..368E} {268, 368}

\bibitem[\protect\citeauthoryear{{Ercolano}, {Barlow}, {Storey}  \&
  {Liu}}{{Ercolano} et~al.}{2003}]{ercolano03}
{Ercolano} B.,  {Barlow} M.~J.,  {Storey} P.~J.,   {Liu} X.-W.,  2003, \mn@doi
  [\mnras] {10.1046/j.1365-8711.2003.06371.x}, \href
  {http://adsabs.harvard.edu/abs/2003MNRAS.340.1136E} {340, 1136}

\bibitem[\protect\citeauthoryear{{Ercolano}, {Barlow}  \& {Storey}}{{Ercolano}
  et~al.}{2005}]{ercolano05}
{Ercolano} B.,  {Barlow} M.~J.,   {Storey} P.~J.,  2005, \mn@doi [\mnras]
  {10.1111/j.1365-2966.2005.09381.x}, \href
  {http://adsabs.harvard.edu/abs/2005MNRAS.362.1038E} {362, 1038}

\bibitem[\protect\citeauthoryear{{Ercolano}, {Young}, {Drake}  \&
  {Raymond}}{{Ercolano} et~al.}{2008}]{ercolano08}
{Ercolano} B.,  {Young} P.~R.,  {Drake} J.~J.,   {Raymond} J.~C.,  2008,
  \mn@doi [\apjs] {10.1086/524378}, \href
  {http://adsabs.harvard.edu/abs/2008ApJS..175..534E} {175, 534}

\bibitem[\protect\citeauthoryear{{Fossati} et~al.,}{{Fossati}
  et~al.}{2010}]{fossati10}
{Fossati} L.,  et~al., 2010, \mn@doi [\apjl] {10.1088/2041-8205/714/2/L222},
  \href {http://adsabs.harvard.edu/abs/2010ApJ...714L.222F} {714, L222}

\bibitem[\protect\citeauthoryear{{Fossati}, {France}, {Koskinen}, {Juvan}, {Haswell}, \& {Lendl}}{{Fossati} et~al.}{2015}]{fossati16}
{Fossati} L., {France} K., {Koskinen} T., {Juvan} I.~G., {Haswell} C.~A., \& Lendl M., \apj, in press (\href {http://arxiv.org/abs/1512.00552} {arXiv:1512.00552})

\bibitem[\protect\citeauthoryear{{Gu}, {Lin}  \& {Bodenheimer}}{{Gu}
  et~al.}{2003}]{gu03}
{Gu} P.-G.,  {Lin} D.~N.~C.,   {Bodenheimer} P.~H.,  2003, \mn@doi [\apj]
  {10.1086/373920}, \href {http://adsabs.harvard.edu/abs/2003ApJ...588..509G}
  {588, 509}

\bibitem[\protect\citeauthoryear{{Haswell} et~al.,}{{Haswell}
  et~al.}{2012}]{haswell12}
{Haswell} C.~A.,  et~al., 2012, \mn@doi [\apj] {10.1088/0004-637X/760/1/79},
  \href {http://adsabs.harvard.edu/abs/2012ApJ...760...79H} {760, 79}

\bibitem[\protect\citeauthoryear{{Hebb} et~al.,}{{Hebb} et~al.}{2009}]{hebb09}
{Hebb} L.,  et~al., 2009, \mn@doi [\apj] {10.1088/0004-637X/693/2/1920}, \href
  {http://adsabs.harvard.edu/abs/2009ApJ...693.1920H} {693, 1920}

\bibitem[\protect\citeauthoryear{{Hellier} et~al.,}{{Hellier}
  et~al.}{2009}]{hellier09}
{Hellier} C.,  et~al., 2009, \mn@doi [\nat] {10.1038/nature08245}, \href
  {http://adsabs.harvard.edu/abs/2009Natur.460.1098H} {460, 1098}

\bibitem[\protect\citeauthoryear{{Iglesias} \& {Rogers}}{{Iglesias} \&
  {Rogers}}{1996}]{opal96}
{Iglesias} C.~A.,  {Rogers} F.~J.,  1996, \mn@doi [\apj] {10.1086/177381},
  \href {http://adsabs.harvard.edu/abs/1996ApJ...464..943I} {464, 943}

\bibitem[\protect\citeauthoryear{{Lai}, {Helling}  \& {van den Heuvel}}{{Lai}
  et~al.}{2010}]{lai10}
{Lai} D.,  {Helling} C.,   {van den Heuvel} E.~P.~J.,  2010, \mn@doi [\apj]
  {10.1088/0004-637X/721/2/923}, \href
  {http://adsabs.harvard.edu/abs/2010ApJ...721..923L} {721, 923}

\bibitem[\protect\citeauthoryear{{Lamers} \& {Cassinelli}}{{Lamers} \&
  {Cassinelli}}{1999}]{lc99}
{Lamers} H.~J.~G.~L.~M.,  {Cassinelli} J.~P.,  1999, {Introduction to Stellar
  Winds}.
{Cambridge University Press}

\bibitem[\protect\citeauthoryear{{Llama}, {Wood}, {Jardine}, {Vidotto},
  {Helling}, {Fossati}  \& {Haswell}}{{Llama} et~al.}{2011}]{llama11}
{Llama} J.,  {Wood} K.,  {Jardine} M.,  {Vidotto} A.~A.,  {Helling} C.,
  {Fossati} L.,   {Haswell} C.~A.,  2011, \mn@doi [\mnras]
  {10.1111/j.1745-3933.2011.01093.x}, \href
  {http://adsabs.harvard.edu/abs/2011MNRAS.416L..41L} {416, L41}

\bibitem[\protect\citeauthoryear{{Llama}, {Vidotto}, {Jardine}, {Wood}, {Fares}
   \& {Gombosi}}{{Llama} et~al.}{2013}]{llama13}
{Llama} J.,  {Vidotto} A.~A.,  {Jardine} M.,  {Wood} K.,  {Fares} R.,
  {Gombosi} T.~I.,  2013, \mn@doi [\mnras] {10.1093/mnras/stt1725}, \href
  {http://adsabs.harvard.edu/abs/2013MNRAS.436.2179L} {436, 2179}

\bibitem[\protect\citeauthoryear{{Lopez}, {Fortney}  \& {Miller}}{{Lopez}
  et~al.}{2012}]{lopez12}
{Lopez} E.~D.,  {Fortney} J.~J.,   {Miller} N.,  2012, \mn@doi [\apj]
  {10.1088/0004-637X/761/1/59}, \href
  {http://adsabs.harvard.edu/abs/2012ApJ...761...59L} {761, 59}

\bibitem[\protect\citeauthoryear{{Mandel} \& {Agol}}{{Mandel} \&
  {Agol}}{2002}]{ma02}
{Mandel} K.,  {Agol} E.,  2002, \mn@doi [\apjl] {10.1086/345520}, \href
  {http://adsabs.harvard.edu/abs/2002ApJ...580L.171M} {580, L171}

\bibitem[\protect\citeauthoryear{{Matsakos}, {Uribe}  \& {K{\"o}nigl}}{{Matsakos}
  et~al.}{2015}]{matsakos15}
{Matsakos} T.,  {Uribe} A.,   {K{\"o}nigl} A.,  2015, \mn@doi [\aap]
  {10.1051/0004-6361/201425593}, \href
  {http://adsabs.harvard.edu/abs/2015A%26A...578A...6M} {578, A6}

\bibitem[\protect\citeauthoryear{{Nichols} et~al.,}{{Nichols}
  et~al.}{2015}]{nichols15}
{Nichols} J.~D.,  et~al., 2015, \mn@doi [\apj] {10.1088/0004-637X/803/1/9},
  \href {http://adsabs.harvard.edu/abs/2015ApJ...803....9N} {803, 9}

\bibitem[\protect\citeauthoryear{{Owen} \& {Jackson}}{{Owen} \&
  {Jackson}}{2012}]{oj12}
{Owen} J.~E.,  {Jackson} A.~P.,  2012, \mn@doi [\mnras]
  {10.1111/j.1365-2966.2012.21481.x}, \href
  {http://adsabs.harvard.edu/abs/2012MNRAS.425.2931O} {425, 2931}

\bibitem[\protect\citeauthoryear{{Parker}}{{Parker}}{1958}]{parker58}
{Parker} E.~N.,  1958, \mn@doi [\apj] {10.1086/146579}, \href
  {http://adsabs.harvard.edu/abs/1958ApJ...128..664P} {128, 664}

\bibitem[\protect\citeauthoryear{{Rasio}, {Tout}, {Lubow}  \& {Livio}}{{Rasio}
  et~al.}{1996}]{rasio96}
{Rasio} F.~A.,  {Tout} C.~A.,  {Lubow} S.~H.,   {Livio} M.,  1996, \mn@doi
  [\apj] {10.1086/177941}, \href
  {http://adsabs.harvard.edu/abs/1996ApJ...470.1187R} {470, 1187}

\bibitem[\protect\citeauthoryear{{Showman} \& {Guillot}}{{Showman} \&
  {Guillot}}{2002}]{sg02}
{Showman} A.~P.,  {Guillot} T.,  2002, \mn@doi [\aap]
  {10.1051/0004-6361:20020101}, \href
  {http://adsabs.harvard.edu/abs/2002A%26A...385..166S} {385, 166}

\bibitem[\protect\citeauthoryear{{Stone} \& {Norman}}{{Stone} \&
  {Norman}}{1992}]{sn92}
{Stone} J.~M.,  {Norman} M.~L.,  1992, \mn@doi [\apjs] {10.1086/191680}, \href
  {http://adsabs.harvard.edu/abs/1992ApJS...80..753S} {80, 753}

\bibitem[\protect\citeauthoryear{{Sutherland} \& {Dopita}}{{Sutherland} \&
  {Dopita}}{1993}]{sd93}
{Sutherland} R.~S.,  {Dopita} M.~A.,  1993, \mn@doi [\apjs] {10.1086/191823},
  \href {http://adsabs.harvard.edu/abs/1993ApJS...88..253S} {88, 253}

\bibitem[\protect\citeauthoryear{{Torres}, {Winn}  \& {Holman}}{{Torres}
  et~al.}{2008}]{torres08}
{Torres} G.,  {Winn} J.~N.,   {Holman} M.~J.,  2008, \mn@doi [\apj]
  {10.1086/529429}, \href {http://adsabs.harvard.edu/abs/2008ApJ...677.1324T}
  {677, 1324}

\bibitem[\protect\citeauthoryear{{Triaud} et~al.,}{{Triaud}
  et~al.}{2010}]{triaud10}
{Triaud} A.~H.~M.~J.,  et~al., 2010, \mn@doi [\aap]
  {10.1051/0004-6361/201014525}, \href
  {http://adsabs.harvard.edu/abs/2010A%26A...524A..25T} {524, A25}

\bibitem[\protect\citeauthoryear{{Tripathi}, {Kratter}, {Murray-Clay}  \&
  {Krumholz}}{{Tripathi} et~al.}{2015}]{tripathi15}
{Tripathi} A.,  {Kratter} K.~M.,  {Murray-Clay} R.~A.,   {Krumholz} M.~R.,
  2015, \mn@doi [\apj] {10.1088/0004-637X/808/2/173}, \href
  {http://adsabs.harvard.edu/abs/2015ApJ...808..173T} {808, 173}

\bibitem[\protect\citeauthoryear{{Vidal-Madjar} et~al.,}{{Vidal-Madjar}
  et~al.}{2013}]{v-m13}
{Vidal-Madjar} A.,  et~al., 2013, \mn@doi [\aap] {10.1051/0004-6361/201322234},
  \href {http://adsabs.harvard.edu/abs/2013A%26A...560A..54V} {560, A54}

\bibitem[\protect\citeauthoryear{{Vidotto}, {Jardine}  \& {Helling}}{{Vidotto}
  et~al.}{2010}]{vidotto10}
{Vidotto} A.~A.,  {Jardine} M.,   {Helling} C.,  2010, \mn@doi [\apjl]
  {10.1088/2041-8205/722/2/L168}, \href
  {http://adsabs.harvard.edu/abs/2010ApJ...722L.168V} {722, L168}

\bibitem[\protect\citeauthoryear{{Vidotto}, {Jardine}  \& {Helling}}{{Vidotto}
  et~al.}{2011}]{vidotto11}
{Vidotto} A.~A.,  {Jardine} M.,   {Helling} C.,  2011, \mn@doi [\mnras]
  {10.1111/j.1745-3933.2010.00991.x}, \href
  {http://adsabs.harvard.edu/abs/2011MNRAS.411L..46V} {411, L46}

\makeatother
\end{thebibliography}


\appendix

\section{Planet-centred accelerations in polar co-ordinates}\label{sec:appendix}
The accelerations due to the planet are evaluated in polar co-ordinates as follows.  The acceleration due to the gravitational and magnetic potential of a planet at position $\mathbf r_{\mathrm p}$=$(r_{\mathrm p},\phi_{\mathrm p})$ at an arbitrary position $\mathbf r$=$(r,\phi)$ is (Equation \ref{eq:a}):
\begin{equation}\label{eq:a_app}
\mathbf a = \left( \frac{C_{\mathrm B}}{|\mathbf r-\mathbf r_{\mathrm p}|^8} - \frac{GM_{\mathrm p}}{|\mathbf r-\mathbf r_{\mathrm p}|^3} \right) (\mathbf r-\mathbf r_{\mathrm p}) \, .
\end{equation}
The relative position vector can be written as
\begin{equation}
\mathbf r-\mathbf r_{\mathrm p} = (r \cos \phi - r_{\mathrm p} \cos \phi_{\mathrm p})\hat{\mathbf x} + (r \sin \phi - r_{\mathrm p} \sin \phi_{\mathrm p})\hat{\mathbf y} \, ,
\end{equation}
where $\hat{\mathbf x}$ \& $\hat{\mathbf y}$ are the usual Cartesian unit vectors, and its magnitude is therefore
\begin{equation}
|\mathbf r-\mathbf r_{\mathrm p}| = \left( r^2 + r_{\mathrm p}^2 - 2 r r_{\mathrm p} \cos (\phi - \phi_{\mathrm p})  \right)^{1/2} \, .
\end{equation}
We write equation (\ref{eq:a_app}) as
\begin{equation}\label{eq:norm}
\mathbf a = A (\mathbf r-\mathbf r_{\mathrm p}) \, ,
\end{equation}
where the normalisation factor $A$(=$a_{\mathrm B}$$+$$a_{\mathrm g}$) is given by
\begin{equation}
A = \frac{C_{\mathrm B}}{|\mathbf r-\mathbf r_1|^8} - \frac{GM_{\mathrm p}}{|\mathbf r-\mathbf r_1|^3} \, .
\end{equation}
To implement these accelerations in polar co-ordinates we must express $\mathbf a$ as a vector $(a_r,a_{\phi})$.  This is done by taking the scalar products of $\mathbf a$ with the unit vectors in the $r$ \& $\phi$ directions, respectively.  The unit vectors here are
\begin{equation}\label{eq:rhat}
\hat{\mathbf r} = \cos \phi \hat{\mathbf x} + \sin\phi \hat{\mathbf y}
\end{equation}
\begin{equation}\label{eq:phihat}
\hat{\mathbf \phi} = -\sin \phi \hat{\mathbf x} + \cos\phi \hat{\mathbf y}
\end{equation}
We then take scalar products of equation (\ref{eq:norm}) with equations (\ref{eq:rhat}) \& (\ref{eq:phihat}) and rearrange to find
\begin{equation}
a_r = A \left(r - r_{\mathrm p} \cos (\phi - \phi_{\mathrm p})\right)
\end{equation}
\begin{equation}
a_{\phi} = A \left( r_{\mathrm p} \sin(\phi - \phi_{\mathrm p})\right)
\end{equation}
These two terms are then added as explicit accelerations in the source step of the {\sc zeus-2d} code. 

\label{lastpage}

\end{document}